\documentclass[aps,amsmath,amssymb,superscriptaddress,longbibliography,preprint,floatfix]{revtex4-2}

\usepackage{acronym}
\usepackage{bm}
\usepackage{color}
\usepackage{derivative}
\usepackage{graphicx}
\usepackage[colorlinks,linkcolor=blue,urlcolor=blue,citecolor=blue,bookmarks=false]{hyperref}
\usepackage[version=4]{mhchem}
\usepackage{siunitx}
\usepackage{physics2}
\usephysicsmodule{ab, braket}

\makeatletter
\let\MYcaption\@makecaption
\makeatother
\usepackage{subcaption}
\makeatletter
\let\@makecaption\MYcaption
\makeatother

\newacro{STM}[STM]{scanning tunneling microscope}
\newacro{HOPG}[HOPG]{highly oriented pyrolytic graphite}
\newacro{QSL}[QSL]{quantum spin liquid}
\newacro{2D}[2D]{two-dimensional}

\graphicspath{{./fig}}

\begin{document}

\title{Imaging quantum interference in a monolayer Kitaev quantum spin liquid candidate}

\newcommand{\Kyoto}{Department of Physics, Kyoto University, Kyoto 606-8502 Japan}
\newcommand{\Tokyo}{Department of Advanced Materials Science, University of Tokyo, Kashiwa, Chiba 277-8561, Japan}
\newcommand{\CEMS}{RIKEN Center for Emergent Matter Science, Wako, Saitama 351-0198, Japan}
\newcommand{\Osaka}{Department of Materials Engineering Science, Osaka University, Toyonaka 560-8531, Japan}
\newcommand{\Gakushuin}{Department of Physics, Gakushuin University, Mejiro, Tokyo, 171-8588, Japan}
\author{Y.~Kohsaka} \email{kohsaka.yuhki.3j@kyoto-u.ac.jp} \affiliation{\Kyoto}
\author{S.~Akutagawa} \affiliation{\Kyoto}
\author{S.~Omachi} \affiliation{\Kyoto}
\author{Y.~Iwamichi} \affiliation{\Kyoto}
\author{T.~Ono} \affiliation{\Kyoto} \affiliation{\Tokyo}
\author{I.~Tanaka} \affiliation{\Kyoto}
\author{S.~Tateishi} \affiliation{\Kyoto}
\author{H.~Murayama} \affiliation{\Kyoto} \affiliation{\CEMS}
\author{S.~Suetsugu} \affiliation{\Kyoto}
\author{K.~Hashimoto} \affiliation{\Tokyo}
\author{T.~Shibauchi} \affiliation{\Tokyo}
\author{M.~O.~Takahashi} \affiliation{\Osaka}
\author{S.~Nikolaev} \affiliation{\Osaka}
\author{T.~Mizushima} \affiliation{\Osaka}
\author{S.~Fujimoto} \affiliation{\Osaka}
\author{T.~Terashima} \affiliation{\Kyoto}
\author{T.~Asaba} \affiliation{\Kyoto}
\author{Y.~Kasahara} \affiliation{\Kyoto}
\author{Y.~Matsuda} \email{matsuda@scphys.kyoto-u.ac.jp} \affiliation{\Kyoto}

\date{\today}

\begin{abstract}
	Single atomic defects are prominent windows to look into host quantum states because collective responses from the host states emerge as localized states around the defects.
	Friedel oscillations and Kondo clouds in Fermi liquids are quintessential examples.
	However, the situation is quite different for \ac{QSL}, an exotic state of matter with fractionalized quasiparticles and topological order arising from a profound impact of quantum entanglement.
	Elucidating the underlying local electronic property has been challenging due to the charge neutrality of fractionalized quasiparticles and the insulating nature of \ac{QSL}s.
	Here, using spectroscopic-imaging scanning tunneling microscopy, we report atomically resolved images of monolayer \ce{\alpha-RuCl3}, the most promising Kitaev \ac{QSL} candidate, on metallic substrates.
	We find quantum interference in the insulator manifesting as incommensurate and decaying spatial oscillations of the local density of states around defects with a characteristic bias dependence.
	The oscillation differs from any known spatial structures in its nature and does not exist in other Mott insulators, implying it is an exotic oscillation involved with excitations unique to \ce{\alpha-RuCl3}.
	Numerical simulations suggest that the observed oscillation can be reproduced by assuming that itinerant Majorana fermions of Kitaev \ac{QSL} are scattered across the Majorana Fermi surface.
	The oscillation provides a new approach to exploring Kitaev \ac{QSL}s through the local response against defects like Friedel oscillations in metals.
\end{abstract}

\maketitle

\section{Introduction}
Uniform electronic states of matter rearrange themselves in response to defects, forming characteristic spatial structures.
The local electronic structure around defects is thus a fundamental fingerprint reflecting low-energy excitations of the host state.
Eminent examples are screening phenomena of metals: Friedel oscillations for charged defects and Kondo clouds for magnetic defects~\cite{Friedel52,Kondo64}.
Advances in techniques of scanning tunneling microscopy allow us to directly image such defect states not only in metals but also in various quantum materials at atomic resolution unavailable by other means~\cite{Pan00,Kitchen06,Ji08}.

Utilizing defect states as \textit{in-situ} probes is envisioned to search for \acf{QSL}, a highly entangled quantum-disordered state of insulating frustrated magnets~\cite{Savary16,Willans10,Vojta16,Das16,Wang21,Kolezhuk06,Ribeiro11,Mross11}.
Depending on the symmetry of the system, several types of \ac{QSL}s and accompanying fractionalized quasiparticles are predicted~\cite{Savary16}.
Among the \ac{QSL}s is the Kitaev \ac{QSL}, which has sparked an explosion of research because of Majorana fermions and non-abelian anyons resulting from the fractionalization of the quantum spin~\cite{Savary16,Takagi19,Motome20,Kitaev06}.
The Kitaev model formulates localized $s = 1/2$ spins on a \ac{2D} honeycomb lattice interacting through bond-dependent Ising couplings.
Noteworthy is that it possesses an exactly solvable ground state, from which Majorana fermions naturally emerge.
This aspect is distinct from the unresolved ground states of triangular and kagome \ac{QSL} candidate systems.
Following the seminal proposal to embody the Kitaev model~\cite{Jackeli09}, a spin-orbit Mott insulator \ce{\alpha-RuCl3} was suggested as a promising candidate~\cite{Plumb14,Johnson15,Kim15,Cao16}.
Since then, a growing body of evidence has been accumulated to indicate the presence of Majorana fermions at low energies in this compound by measurements of Raman scattering, inelastic neutron scattering, specific heat, and thermal Hall effect~\cite{Sandilands15,Nasu16,Banerjee16,Banerjee17,Do17,Widmann19,Tanaka22,Kasahara18,Yamashita20,Yokoi21,Bruin22}.
Despite such extensive studies, there is still room for debate on whether the Kitaev \ac{QSL} is realized in \ce{\alpha-RuCl3}~\cite{Czajka22}.

\begin{figure}[t]
	\centering
	\begin{subcaptiongroup}
		\includegraphics{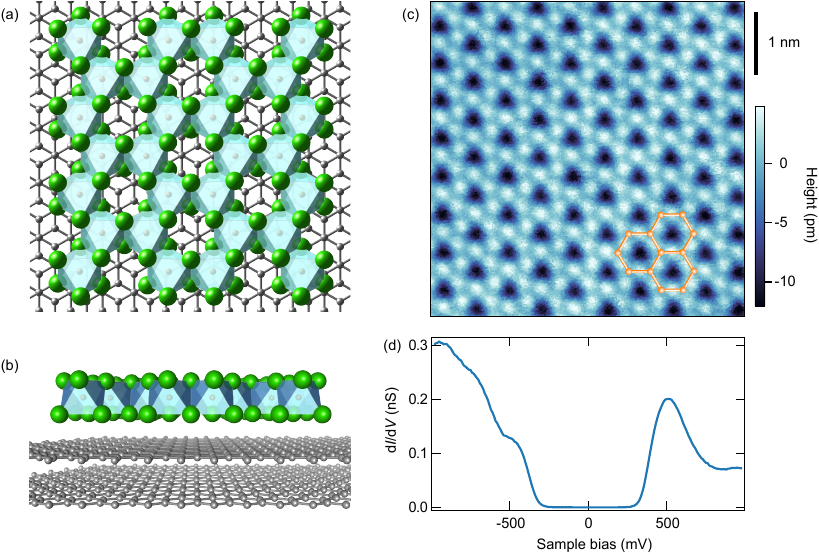}
		\phantomcaption\label{fig:crystal_structure:top}
		\phantomcaption\label{fig:crystal_structure:side}
		\phantomcaption\label{fig:successful_fabrication:topo}
		\phantomcaption\label{fig:successful_fabrication:dIdV}
	\end{subcaptiongroup}
	\caption{
		A monolayer \ce{\alpha-RuCl3} film fabricated on a graphite substrate.
		(a), (b) Illustrations of a monolayer \ce{\alpha-RuCl3} film grown on a graphite substrate, drawn using VESTA~\cite{Momma11}.
		\subref{fig:crystal_structure:top} and \subref{fig:crystal_structure:side} are seen from $(001)$ and $(100)$ directions of the monolayer \ce{\alpha-RuCl3} film, respectively.
		The relative angle between the \ce{\alpha-RuCl3} lattice and the graphite lattice drawn in the illustrations differs from the actual angles.
		(c) A topographic image of a monolayer \ce{\alpha-RuCl3} film.
		The setpoint condition is \SI[explicit-sign=+]{1}{V} and \SI{0.2}{nA}.
		The overlaid illustration depicts the position of the Ru honeycomb.
		(d) A typical conductance spectrum.
		The setpoint condition is \SI[explicit-sign=+]{0.98}{V} and \SI{0.2}{nA}.
	}\label{fig:successful_fabrication}
\end{figure}

The quest for \ac{QSL}s, including the Kitaev \ac{QSL}, has been driven by spatially averaged probes, as exemplified above.
Consequently, the experimental data and their interpretations have often been influenced by undesirable complexities due to structural disorders such as stacking faults and antisite defects~\cite{Cao16,Zhu17}.
Using spatially resolved probes on thin films is a reasonable way to circumvent these issues.
Additionally, for \ce{\alpha-RuCl3}, thin film samples can serve as an alternative to applying a horizontal magnetic field to suppress the zigzag antiferromagnetism, which is a three-dimensional order~\cite{Balz21}.
Furthermore, beyond the general context to investigate defect states described above, local electronic probes have been theoretically proposed to detect and control fractional magnetic excitations~\cite{Pereira20,Feldmeier20,Konig20,Udagawa21,Jang21,Bauer23}.
These previous studies underscore the need for scanning tunneling microscopy of monolayer thin films.
Recently, monolayer 1$T$-\ce{TaSe2}, a candidate for another \ac{QSL} in a \ac{2D} triangular lattice Mott insulator, has been examined by an \ac{STM} ~\cite{Ruan21,Chen22}.
These pioneering works have suggested that the low-energy magnetic excitations could be detected experimentally at higher energies outside the Mott gap by measuring the tunneling electrons recombined from the fractionalized quasiparticles, called spinons.
Raman scattering study of exfoliated monolayer films of \ce{\alpha-RuCl3} uncovered enhanced frustrated magnetic interactions and the unusual magnetic continuum indicating proximate \ac{QSL} in the \ac{2D} samples~\cite{Du18}.
However, atomically-resolved \ac{STM} studies revealed that exfoliated samples can be deformed and exhibit highly inhomogeneous spectra~\cite{Zheng23,Zheng24}.
These results highlight the need for STM measurements on thin films fabricated in a more controlled manner.
We therefore fabricated monolayer \ce{\alpha-RuCl3} films on \ac{HOPG} substrates by pulsed laser deposition (Figs.~\ref{fig:crystal_structure:top} and \ref{fig:crystal_structure:side}) and conducted the electronic imaging study using an \ac{STM}.
(See \hyperref[sec:Methods]{Methods} for details.)

\section{Results}
We first inspect fabricated films.
Figure~\ref{fig:successful_fabrication:topo} shows an atomic-resolution topographic image of our film.
Regularly-arranged circular protrusions form a Kagome-like lattice, replicating the previous study of monolayer \ce{\alpha-RuCl3}~\cite{Wang22}.
The protrusions are primarily ascribed to derive from the topmost Cl $p$ orbitals, with the Ru site residing at the center of three protrusions.
Due to the hybridization of Cl $p$ and Ru $d$ orbitals, the center of the protrusion slightly shifts from the Cl site towards the Ru site, resulting in the Kagome-like lattice~\cite{Wang22}.
A Ru-honeycomb encloses a dark-colored hollow site, as shown in the overlaid illustration of Fig.~\ref{fig:successful_fabrication:topo}.
The conductance spectrum exemplified in Fig.~\ref{fig:successful_fabrication:dIdV} is also similar to the previous study.
It shows an energy gap of about \SI{0.6}{eV} with the Fermi energy in the middle of the gap, indicating that the sample is an insulator.
The apparent film thickness is also very close to the value in Ref.~\cite{Wang22}.
(See \hyperref[sec:film_thickness]{Supplementary Material \ref{sec:film_thickness}} for details.)
From these close similarities with the previous study, we identify that the films are monolayer \ce{\alpha-RuCl3}.

Our spectra are strikingly different from the metallic spectra of the exfoliated films~\cite{Yang22,Zheng23,Zheng24}, indicating that our samples are unaffected by the lattice deformation.
Furthermore, the Fermi energy in the middle of the gap does not align with sizable electron transfer from the HOPG substrate observed in some experiments~\cite{Mashhadi19,Zhou19,Rizzo20}.
We note that the samples in these experiments are also prepared by exfoliation.
Practically, the properties of insulators or metals can vary depending on the fabrication methods: deposition and epitaxy, or exfoliation.
In our case, the weak coupling between the \ce{\alpha-RuCl3} film and the HOPG substrate, inferred from the apparent film thickness and the random orientation (described below), may be preventing electron transfer and thus keeping the sample insulating.

\begin{figure}[ht]
	\begin{subcaptiongroup}
		\includegraphics{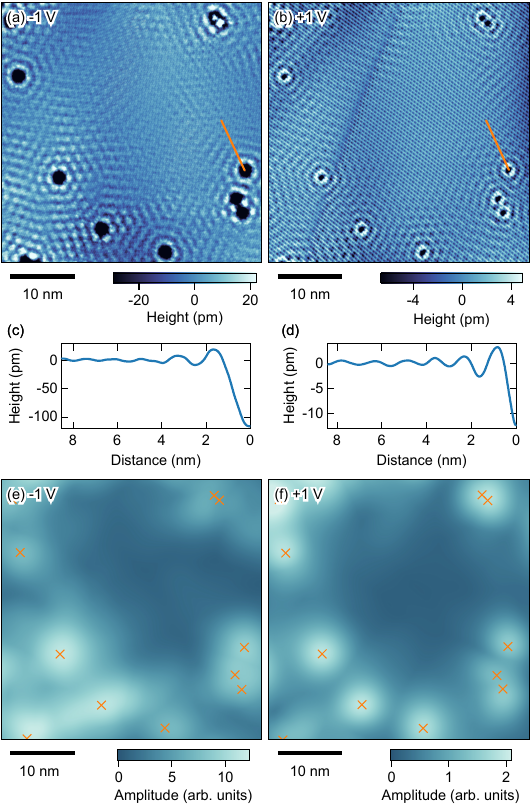}
		\phantomcaption\label{fig:oscillation:topo_-1V}
		\phantomcaption\label{fig:oscillation:topo_+1V}
		\phantomcaption\label{fig:oscillation:decay_-1V}
		\phantomcaption\label{fig:oscillation:decay_+1V}
		\phantomcaption\label{fig:oscillation:LAM_-1V}
		\phantomcaption\label{fig:oscillation:LAM_+1V}
	\end{subcaptiongroup}
	\caption{
		Spatial oscillation of the local density of states decaying away from defects.
		(a), (b) Topographic images of an \ce{\alpha-RuCl3} monolayer film taken at \SI[explicit-sign=+]{-1}{V} and \SI{0.25}{nA} for \subref{fig:oscillation:topo_-1V} and \SI[explicit-sign=+]{1}{V} and \SI{0.5}{nA} for \subref{fig:oscillation:topo_+1V}.
		The orange lines denote the positions of the line profiles in \subref{fig:oscillation:decay_-1V} and \subref{fig:oscillation:decay_+1V}.
		(c), (d) The line profiles obtained from the low-pass filtered images shown in Fig.~\ref{fig:S_filter} along the trajectories shown in \subref{fig:oscillation:topo_-1V} and \subref{fig:oscillation:topo_+1V}, respectively.
		(e), (f) The local amplitude maps of the oscillation in the same field of view as \subref{fig:oscillation:topo_-1V} and \subref{fig:oscillation:topo_+1V}, respectively.
		The orange markers are placed at the defect centers.
		The calculation of the local amplitude map is described in \hyperref[sec:local_amplitude_map]{Supplementary Material \ref{sec:local_amplitude_map}}.
	}\label{fig:oscillation_decay}
\end{figure}

We next focus on the most peculiar feature we have observed.
Figures~\ref{fig:oscillation:topo_+1V} and \ref{fig:oscillation:topo_-1V} are topographic images taken at \SI{\pm 1}{V} in the same field of view.
We find concentric oscillatory patterns around defects.
High-resolution measurements allow us to identify several types of defects (Fig.~\ref{fig:S_defects}), showing that the oscillation appears independently of the defect sites.
The oscillation is several times larger in amplitude at \SI{-1}{V} than at \SI[explicit-sign=+]{1}{V} and decays away from the defects, as highlighted in Figs.~\ref{fig:oscillation:decay_-1V} and \ref{fig:oscillation:decay_+1V} (see also Fig.~\ref{fig:S_filter}).
The local amplitude maps of the oscillation, as depicted in Figs.~\ref{fig:oscillation:LAM_-1V} and \ref{fig:oscillation:LAM_+1V}, further demonstrate that the oscillation is localized around the defects and diminishes with increasing distance from them.
The decay means that the oscillation is not a moir\'{e} pattern between the monolayer \ce{\alpha-RuCl3} and the substrate.

\begin{figure}[ht]
	\begin{subcaptiongroup}
		\includegraphics{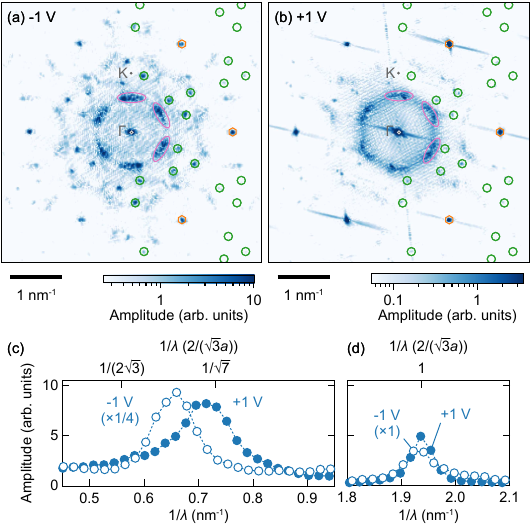}
		\phantomcaption\label{fig:oscillation:FT_-1V}
		\phantomcaption\label{fig:oscillation:FT_+1V}
		\phantomcaption\label{fig:oscillation:azimuthal_average_incomm}
		\phantomcaption\label{fig:oscillation:azimuthal_average_Bragg}
	\end{subcaptiongroup}
	\caption{
		(a), (b) Fourier transforms of images including \subref{fig:oscillation:topo_-1V} and \subref{fig:oscillation:topo_+1V}, respectively.
		The images are \qtyproduct{54 x 43}{nm} for \subref{fig:oscillation:FT_-1V} and \qtyproduct{54 x 54}{nm} for \subref{fig:oscillation:FT_+1V}.
		The pink ovals indicate the peaks corresponding to the oscillation decaying away from the defects.
		The orange hexagons and the green circles denote the positions of the \ce{\alpha-RuCl3} lattice peaks and the satellite peaks, respectively.
		The satellite peak positions are calculated as described in \hyperref[sec:satellite_peaks]{Supplementary Material \ref{sec:satellite_peaks}} with an angle of \ang{31}.
		The markers are shown only in the right half of each panel.
		The defects in \subref{fig:oscillation:topo_-1V} are masked before calculating \subref{fig:oscillation:FT_-1V} to suppress large intensity around the origin, as shown in Fig.~\ref{fig:S_mask}.
		(c), (d) The azimuthal averages of \subref{fig:oscillation:FT_-1V} and \subref{fig:oscillation:FT_+1V}.
		The top axis is shown in units of the inverse of the lattice constant.
		The vertical axis range is common to \subref{fig:oscillation:azimuthal_average_incomm} and \subref{fig:oscillation:azimuthal_average_Bragg}.
	}\label{fig:oscillation_FT}
\end{figure}

To analyze the wavevectors of oscillation, we calculated the Fourier transform, as shown in Figs.~\ref{fig:oscillation:FT_-1V} and \ref{fig:oscillation:FT_+1V}.
We suppressed long-wavelength features from the defects for clarity, as demonstrated in Fig.~\ref{fig:S_mask}.
Strong peaks corresponding to the oscillations, indicated by the pink ovals, are found roughly in the $\Gamma$--K direction.
The wavenumbers of the oscillation are incommensurate.
(See also \hyperref[sec:possible_commensurate_structure]{Supplementary Material~\ref{sec:possible_commensurate_structure}} for more details.)
More critically, they differ between the \SI{\pm 1}{V} images, as shown in Fig.~\ref{fig:oscillation:azimuthal_average_incomm}.
In contrast, the wavevectors of the lattice peaks are independent of the bias voltage, as shown in Fig.~\ref{fig:oscillation:azimuthal_average_Bragg}.
Similarly, the peak positions of charge density waves do not depend on the bias voltage for the same reason~\cite{Ruan21}.
Namely, the wavenumbers different between \SI{\pm 1}{V} indicate that the origin of the incommensurate oscillation is electronic rather than structural.
Consequently, we exclude phenomena involving lattice distortion, such as charge density waves and local strains, as the origin of the oscillation.
In addition, due to the decaying feature and the different wavenumbers, the oscillation is distinguished from the long-wavelength super-modulations in monolayer $1T$-\ce{TaSe2}~\cite{Ruan21}.
Interestingly, a similar pattern has been independently observed by another group~\cite{Qiu24}.
There are both similarities and differences: the similarities include an apparent resemblance, especially at negative biases, and different wavenumbers between the polarities; the differences include the presence of a decaying feature, incommensurate oscillations, and no apparent rotational symmetry breaking in our case.

The Fourier transforms also show many bias-independent peaks (the green circles) besides the lattice peaks (the orange hexagons).
These are satellite peaks generated by the \ce{\alpha-RuCl3} lattice and the substrate \ac{HOPG} lattice with a certain angle.
(How to calculate the peak positions and the angle is described in \hyperref[sec:satellite_peaks]{Supplementary Material~\ref{sec:satellite_peaks}}.)
We also found a monolayer \ce{\alpha-RuCl3} film with a different angle and observed the same oscillations (Fig.~\ref{fig:S_25deg}).
The insensitivity to the relative angles demonstrates that the oscillation is irrelevant to coupling with the substrate.
Moreover, electron tunneling directly from the substrate is negligibly small, as evidenced by the zero conductance in the insulating gap (Fig.~\ref{fig:successful_fabrication:dIdV}).
Therefore, the oscillation is inherent to the monolayer \ce{\alpha-RuCl3} and occurs in the monolayer \ce{\alpha-RuCl3}, neither in the substrate nor at the interface between the monolayer \ce{\alpha-RuCl3} and the substrate.

\begin{figure}[t]
	\centering
	\begin{subcaptiongroup}
		\includegraphics{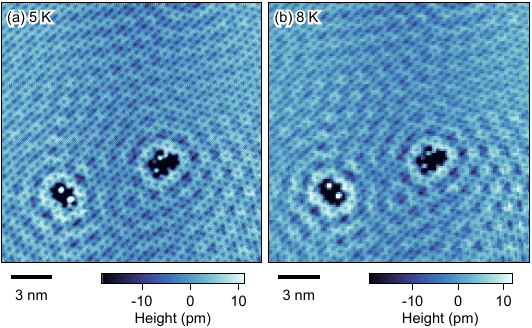}
		\phantomcaption\label{fig:8K:topo_5K}
		\phantomcaption\label{fig:8K:topo_8K}
	\end{subcaptiongroup}
	\caption{
		Comparison of topographic images taken at \SI{5}{K} and \SI{8}{K}.
		Both images were taken in the same field of view of \qtyproduct{19 x 19}{nm} and at a setpoint condition of \SI[explicit-sign=+]{0.98}{V} and \SI{0.1}{nA}.
	}\label{fig:8K}
\end{figure}

Since the oscillation is electronic in origin and occurs in \ce{\alpha-RuCl3}, one may wonder if the zigzag antiferromagnetic order found in the bulk \ce{\alpha-RuCl3} is relevant to the oscillation.
The fact that the oscillation patterns are not unidirectional but approximate hexagonal shapes indicates that they are unrelated to the zigzag antiferromagnetic order.
Furthermore, the temperature dependence of the pattern also supports the irrelevance.
In the presence of the Kitaev interaction, the zigzag antiferromagnetic order arising from non-Kitaev interactions is indeed allowed even in \ac{2D} without being forbidden by the Mermin--Wagner theorem because the Kitaev interaction has $Z_2$ symmetry~\cite{Price13}.
However, even if it exists, the N\'{e}el temperature is expected to be lower in \ac{2D} films than in the bulk since the zigzag antiferromagnetic order is three-dimensional~\cite{Balz21}.
Moreover, the Imry--Ma argument indicates that long-range magnetic orders with $Z_2$ symmetry are destroyed in \ac{2D} by infinitesimally weak disorders~\cite{Imry75}.
Therefore, we presume that the zigzag antiferromagnetic order is absent at \SI{8}{K}, higher than the N\'{e}el temperature of \SI{7}{K} in the bulk.
As shown in Fig.~\ref{fig:8K}, a topographic image taken at \SI{8}{K} shows no discernable difference from one at \SI{5}{K}, indicating that the oscillation has nothing to do with the zigzag antiferromagnetic order.

\begin{figure}[t]
	\centering
	\begin{subcaptiongroup}
		\includegraphics{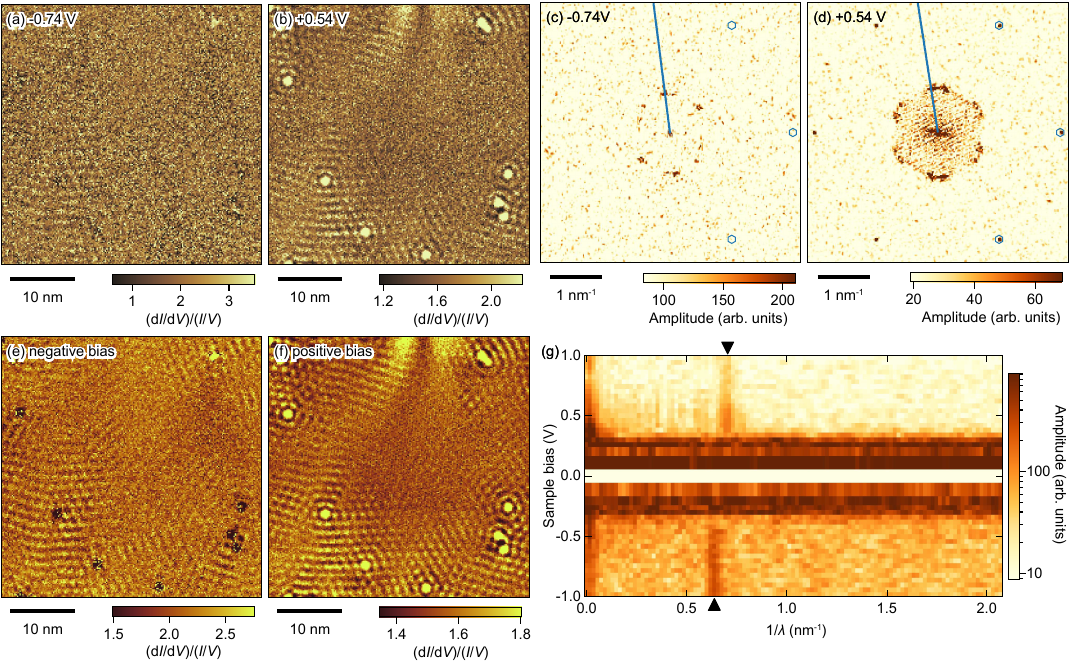}
		\phantomcaption\label{fig:map:-0.74V}
		\phantomcaption\label{fig:map:+0.54V}
		\phantomcaption\label{fig:map:-0.74V_FT}
		\phantomcaption\label{fig:map:+0.54V_FT}
		\phantomcaption\label{fig:map:avg_negative}
		\phantomcaption\label{fig:map:avg_positive}
		\phantomcaption\label{fig:map:dispersion}
	\end{subcaptiongroup}
	\caption{
		The maps of normalized conductance $(\odv{I}/{V})/(I/V)$.
		(a), (b) The normalized conductance maps taken at \SI{-0.74}{V} and \SI[explicit-sign=+]{0.54}{V}, respectively, in the same field of view of Figs.~\ref{fig:oscillation:topo_-1V} and \ref{fig:oscillation:topo_+1V}.
		(c), (d) Fourier transforms of normalized conductance maps.
		The original images were measured in a \qtyproduct{54 x 53}{nm} field of view, including \subref{fig:map:-0.74V} and \subref{fig:map:+0.54V}, with a setpoint condition of \SI[explicit-sign=+]{0.98}{V} and \SI{0.1}{nA}.
		The blue hexagons in the right half of each panel denote the positions of the lattice peaks.
		(e), (f) The normalized conductance maps averaged between \SI{-1}{V} and \SI{-0.42}{V} for \subref{fig:map:avg_negative} and \SI[explicit-sign=+]{0.42}{V} and \SI[explicit-sign=+]{1}{V} for \subref{fig:map:avg_positive}.
		The field of view is the same as \subref{fig:map:-0.74V} and \subref{fig:map:+0.54V}.
		(g) Dispersion relation along the blue lines in \subref{fig:map:-0.74V_FT} and \subref{fig:map:+0.54V_FT}.
		The triangle markers indicate the bias-semi-independent wavevectors of the oscillatory patterns.
	}\label{fig:map}
\end{figure}

The oscillatory patterns in the \ac{STM} images (and also the conductance maps described later) decaying away from the defects imply quantum interference of fermionic quasiparticles around the defects.
At first glance, the patterns resemble the Friedel oscillations in metals and quasiparticle interference.
Although a straightforward case of Friedel oscillations is ruled out because the monolayer \ce{\alpha-RuCl3} is insulating, as evidenced by the energy gap (Fig.~\ref{fig:successful_fabrication:dIdV}), Friedel oscillations could potentially occur if the tip-induced band bending is sufficient to induce carriers.
In this case, the oscillation period would depend on the density of induced carriers and the bias voltages.
The wavevectors of quasiparticle interference, which reflect the band structure, also vary with the energy.
Therefore, the dispersion relationship of the oscillation is crucial to explore these possibilities.
Figure~\ref{fig:map} displays the result of spectroscopic imaging we have performed to delineate the dispersion relation.
We adopt the normalized conductance $[(\odv{I}/{V})/(I/V)]$ map (Figs.~\ref{fig:map:-0.74V} and \ref{fig:map:-0.74V}) rather than the raw conductance $(\odv{I}/{V})$ map to mitigate the setpoint effect
(Fig.~\ref{fig:S_setpoint_effect}).
Figures~\ref{fig:map:-0.74V_FT} and \ref{fig:map:+0.54V_FT} show Fourier transforms of conductance maps.
Peaks corresponding to the oscillation are observed in the bias range outside the energy gap.
Notably, the wavevectors differ between the polarities but do not change in each polarity, as shown in Fig.~\ref{fig:map:dispersion}.
The wavevector at each polarity is the same as that observed in the corresponding topographic image.
Indeed, the normalized conductance images averaged for the negative and positive bias voltages (Figs.~\ref{fig:map:avg_negative} and \ref{fig:map:avg_positive}) exhibit oscillations similar to those in the topographic images (Figs.~\ref{fig:oscillation:topo_-1V} and \ref{fig:oscillation:topo_+1V}).
We refer to this behavior of the experimental data as semi-independent of the bias voltage.
The bias-independent aspect indicates that the oscillation is neither Friedel oscillations by the induced carriers nor quasiparticle interference.
We note that non-dispersive quasiparticle interference, requiring electron bands to be parallelly shifted, results in vanishing interference intensity due to destructive interference~\cite{Jolie18}.

\section{Discussion}
As mentioned above, the observed patterns differ from the known phenomena producing oscillatory patterns, such as moir\'{e}, charge density waves, Friedel oscillations, quasiparticle interference, and the super-modulation in monolayer $1T$-\ce{TaSe2}~\cite{Ruan21}.
Therefore, we conclude that the observed oscillation is an unprecedented oscillatory phenomenon.
Since the oscillation appears in the energy range of lower and upper Hubbard bands, one could assume that the Hubbard interaction is involved in the oscillation.
However, no other Mott insulators exhibit oscillations around defects~\cite{Ye13,Dai14,Wu22,Wang20}.
Also, one might consider that the oscillation is special to monolayer films.
However, the decaying and bias-semi-independent oscillation is not found in other insulating monolayer films~\cite{Chen19,Ruan21,Chen22,Wang20,Quertite21}.
Therefore, something unique to \ce{\alpha-RuCl3} is likely to be responsible for the oscillation.

\begin{figure}[ht]
	\centering
	\begin{subcaptiongroup}
		\includegraphics{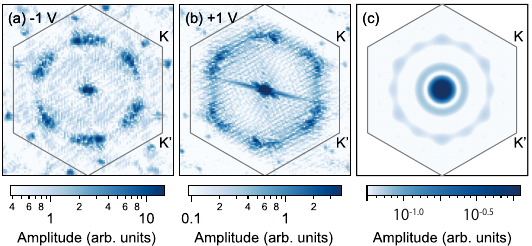}
		\phantomcaption\label{fig:FT_comparison:-1V}
		\phantomcaption\label{fig:FT_comparison:+1V}
		\phantomcaption\label{fig:FT_comparison:simulation}
	\end{subcaptiongroup}
	\caption{
		Comparison between the experimental data and the numerical simulation.
		(a), (b) Fourier transforms of topographic images at \SI{-1}{V} and \SI[explicit-sign=+]{1}{V}, respectively.
		These are the central part of Figs.~\ref{fig:oscillation:FT_-1V} and \ref{fig:oscillation:FT_+1V}.
		The hexagon depicts the first Brillouin zone.
		(c) Fourier transform of the numerical simulation of charge density variation.
		This figure is the same as Fig.~\ref{fig:S_charge_density:bound_3.5_FT}.
	}\label{fig:FT_comparison}
\end{figure}

Given that \ce{\alpha-RuCl3} is a promising candidate for Kitaev \ac{QSL}, we might consider the Kitaev interaction as a possible origin of the oscillation.
Then, two immediate questions arise: How are the spin properties amenable to detection using non-magnetic scanning tips, and what determines the length scale of the incommensurate oscillation?
For the former, if spin--charge separation occurs, the tunneling electrons recombined from spinons and chargons may carry spin information~\cite{Ruan21,Chen22,Mross11}.
However, this process is not the case for the Kitaev \ac{QSL} because fractionalization occurs solely in the spin system.
Instead, we consider a relationship in the Mott insulator that the charge density is tied to the spin correlation function~\cite{Pereira20,Bulaevskii08}.
A spatial texture of the spin correlation function is then reflected in the charge density variation, which is readily imaged as a bias-independent pattern using an \ac{STM} with a non-magnetic scanning tip.
For the length scale of the oscillation, the incommensurability of the oscillation hints at the scattering of itinerant quasiparticles with a characteristic length akin to a Fermi wavelength.
In the Kitaev \ac{QSL}, the spins are fractionalized into itinerant and localized Majorana fermions; the former move around the whole crystal, while the latter form a $Z_2$-vortex called vison~\cite{Kitaev06}.
Thus, itinerant Majorana fermions could play an essential role in the oscillation.
However, for the pure Kitaev model, the scattering vectors of the itinerant Majorana fermions at the Fermi energy are commensurate because the Dirac points of the Majorana band cross the Fermi energy at K and K$^\prime$ points in the Brillouin zone.
Nevertheless, the Majorana Fermi surface with incommensurate Fermi wavenumbers is possibly realized if there are perturbations breaking time-reversal and inversion symmetries that protect the positions of the Dirac points~\cite{Takikawa19,Shiozaki22}.
Postulating that both symmetries are locally broken by the tunneling current injected from the scanning tip, the calculations of charge density variation reproduce the observed incommensurate oscillation, as shown in Figs.~\ref{fig:FT_comparison}.
In this scenario, the slightly different wavenumbers depending on the bias polarities could be attributed to the fact that the correction to the Kitaev interaction due to the bias voltages is asymmetric with respect to the sign of the voltages.
In fact, a negative bias potential imposed by the scanning tip may induce virtual processes with extra holes associated with the Kitaev interaction arising from the Hund coupling between the $j=1/2$ and $j=3/2$ states, which results in the increase of the Kitaev interaction as described by
$\sim \frac{t^2J_\mathrm{H}}{U^2}+J_\mathrm{H}\ab(\frac{t}{U})^2\ab(\frac{t_\mathrm{S}}{U})^2$.
Here, the first term represents the conventional Kitaev interaction, and the second term is an enhancement in the negative bias.
$J_\mathrm{H}$ is the Hund coupling, $U$ is the on-site Coulomb repulsion, $t$ is the hopping amplitude between neighboring sites, and $t_\mathrm{S}$ is the amplitude for the tunneling of the extra holes.
On the other hand, such processes increasing the Kitaev interaction are absent for positive bias voltages.
This prediction is confirmed by model calculations, which show that the difference of the Kitaev interaction between the negative and positive bias voltages is estimated as $(K_\mathrm{negative}-K_\mathrm{positive})/K \sim 0.0543 \sim 0.176$, where $K$ is the Kitaev interaction without the corrections.
The magnitude of the difference varies depending on the choice of parameters.
As the Kitaev interaction increases, the ratio of scalar chirality to Kitaev interaction decreases, leading to smaller wavenumbers, which explains the different wavenumbers shown in Fig.~\ref{fig:map:dispersion}.
(The discussion in this paragraph is detailed in \hyperref[sec:Majorana_scenario]{Supplementary Material~\ref{sec:Majorana_scenario}}.)

While we have suggested the origin of the oscillation as described above, there may be different explanations~\cite{Zhang24} and a more comprehensive understanding is open for future research.
Importantly, however, the decaying, incommensurate, and bias-semi-independent oscillation we found in the insulator manifests a new oscillatory phenomenon.
The unforeseen oscillation represents the atomic-scale response of the quantum state with characteristic length scales.
The absence of such oscillations in other Mott insulators and monolayer films implies that the observed oscillation may serve as a local signifier of Kitaev QSL experimentally elusive.

\section{Methods}\label{sec:Methods}
\subsection{Sample fabrication}
Monolayer \ce{\alpha-RuCl3} films were deposited on \ac{HOPG} substrates by pulsed laser deposition using a yttrium-aluminium-garnet laser (wavelength \SI{1064}{nm}).
The targets were pelletized \ce{\alpha-RuCl3} single crystals grown by chemical vapor transport from commercial \ce{RuCl3} powder.
The chlorine partial pressure and the substrate temperature were optimized at \SI{2000}{Pa} and \SI{430}{\degreeCelsius}.
This condition is essential to grow the $\alpha$ phase separately from the $\beta$ phase~\cite{Asaba23} without mixing the two phases~\cite{Wang22}.
The fabricated thin films were transferred from the deposition chamber to the \ac{STM} chamber without air exposure using a portable ultra-high vacuum chamber.

\subsection{Spectroscopic-imaging scanning tunneling microscopy}
Spectroscopic imaging scanning tunneling microscopy measurements were performed using a low-temperature ultra-high vacuum system (UNISOKU USM 1300).
The scanning tips were mechanically sharpened Pt-Ir wires cleaned by electron-beam heating and conditioned on clean Au(111) surfaces.
All the measurements were carried out at \SI{5}{K} unless otherwise noted.
Topographic images were recorded in the constant-current mode.
Differential conductance spectra were measured using a standard lock-in technique with a modulation amplitude of \SI{20}{meV} at a frequency of \SI{973}{Hz}.
The normalized conductance is obtained by numerical division.
When Fourier transforms are calculated, affine transformations are applied as described in \hyperref[sec:satellite_peaks]{Supplementary Material \ref{sec:satellite_peaks}}, and no symmetrization is used.

\begin{acknowledgements}
The authors thank J. Knolle, E. -G. Moon, Y. Motome, J. Nasu, M. Udagawa, and M. G. Yamada for fruitful discussions.
This work was supported by CREST (No. JPMJCR19T5) from Japan Science and Technology (JST), and Grants-in-Aid for Scientific Research (KAKENHI) (Nos. JP20K03860, JP21H01039, JP22J20066, JP22H00105, JP22H01221, JP23H00089, and JP23K17669) and Grant-in-Aid for Scientific Research on Innovative Areas ``Quantum Liquid Crystals'' (Nos. JP19H05824, JP22H04480) from Japan Society for the Promotion of Science (JSPS).
{M. O. T.} acknowledges support from a JSPS Fellowship for Young Scientists.
\end{acknowledgements}

\bibliography{main}

\clearpage
\setcounter{figure}{0}
\setcounter{section}{0}
\setcounter{subsection}{0}
\makeatletter
\renewcommand{\thefigure}{S\@arabic\c@figure}
\renewcommand{\theequation}{S\@arabic\c@equation}
\makeatother

\section*{Supplementary Material}\label{sec:SI}

\begin{figure}[ht]
	\hspace{-5mm}
	\begin{subcaptiongroup}
		\includegraphics{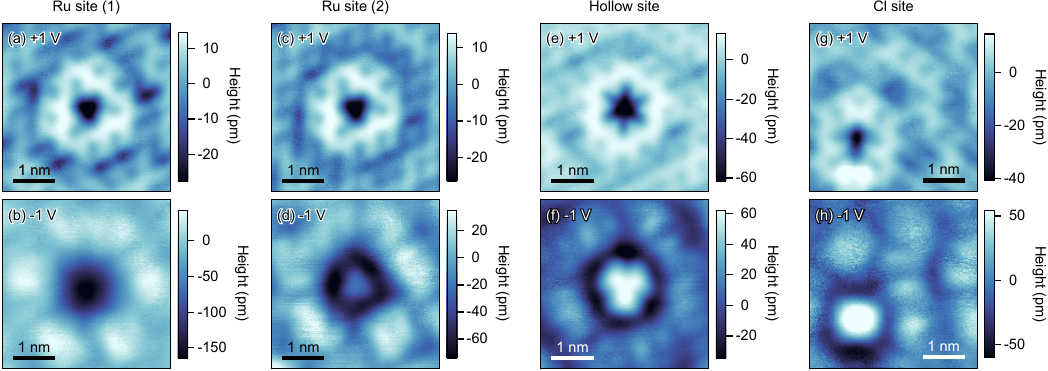}
		\phantomcaption\label{fig:S_defects:Ru1_+1V}
		\phantomcaption\label{fig:S_defects:Ru1_-1V}
		\phantomcaption\label{fig:S_defects:Ru2_+1V}
		\phantomcaption\label{fig:S_defects:Ru2_-1V}
		\phantomcaption\label{fig:S_defects:Hollow_+1V}
		\phantomcaption\label{fig:S_defects:Hollow_-1V}
		\phantomcaption\label{fig:S_defects:Cl_+1V}
		\phantomcaption\label{fig:S_defects:Cl_-1V}
	\end{subcaptiongroup}
	\caption{
		High-resolution topographic images of defects.
		The setpoint voltages are written at the top left corner of each panel.
		The setpoint currents are \SI{0.5}{nA} for \subref{fig:S_defects:Ru1_+1V}, \subref{fig:S_defects:Ru2_+1V}, and \subref{fig:S_defects:Hollow_+1V}, \SI{0.25}{nA} for \subref{fig:S_defects:Ru1_-1V}, \subref{fig:S_defects:Ru2_-1V}, and \subref{fig:S_defects:Hollow_-1V}, \SI{0.2}{nA} for \subref{fig:S_defects:Cl_+1V}, and \SI{20}{pA} for \subref{fig:S_defects:Cl_-1V}.
		The defect sites, written at the top of each column, are identified from the positions relative to the topmost atoms.
		The difference between Ru (1) and (2) is not the Ru sublattices but the apparent height at the center of defects at \SI{-1}{V}.
		The Ru (1) and (2) are indeed on the same Ru sublattice, as indicated by the orientation of the triangular patterns observed around the defects at \SI[explicit-sign=+]{1}{V}.
		The orientation is opposite for the other sublattice.
		Note that the identities of defects (elements of impurities or vacancies) can not be determined solely from the \ac{STM} measurements.
	}\label{fig:S_defects}
\end{figure}

\begin{figure}[ht]
	\centering
	\begin{subcaptiongroup}
		\includegraphics{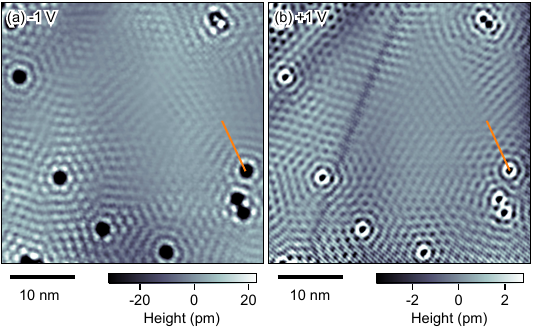}
		\phantomcaption\label{fig:S_filter:topo_-1V}
		\phantomcaption\label{fig:S_filter:topo_+1V}
	\end{subcaptiongroup}
	\caption{
		Low-pass filtered topographic images.
		The lattice peaks (the orange hexagons in Figs.~\ref{fig:oscillation:topo_-1V} and \ref{fig:oscillation:topo_+1V}) and the satellite peaks (the green circles in Figs.~\ref{fig:oscillation:topo_-1V} and \ref{fig:oscillation:topo_+1V}) are also filtered out.
		\subref{fig:S_filter:topo_-1V} and \subref{fig:S_filter:topo_+1V} are filtered images of Figs.~\ref{fig:oscillation:topo_-1V} and \ref{fig:oscillation:topo_+1V}, respectively.
		The line profiles shown in Figs.~\ref{fig:oscillation:decay_-1V} and \ref{fig:oscillation:decay_+1V} are obtained along the orange lines in \subref{fig:S_filter:topo_-1V} and \subref{fig:S_filter:topo_+1V}.
	}\label{fig:S_filter}
\end{figure}

\begin{figure}[ht]
	\centering
	\begin{subcaptiongroup}
		\includegraphics{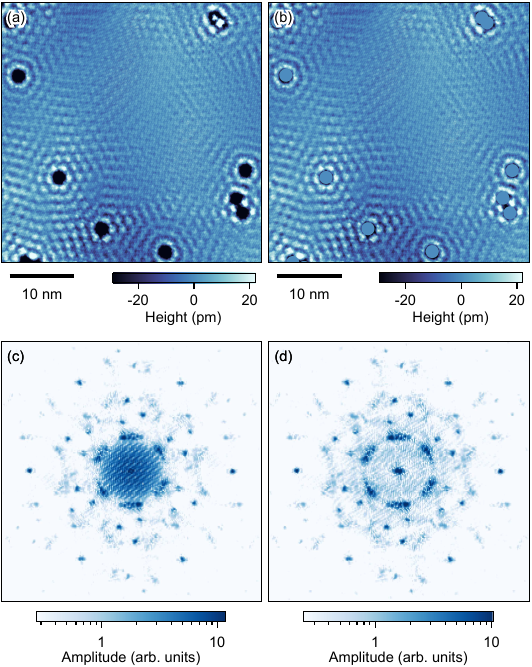}
		\phantomcaption\label{fig:S_mask:topo}
		\phantomcaption\label{fig:S_mask:topo_masked}
		\phantomcaption\label{fig:S_mask:FT}
		\phantomcaption\label{fig:S_mask:FT_masked}
	\end{subcaptiongroup}
	\caption{
		The effect of masking defects before calculating the Fourier transform.
		(a) The same image as Fig.~\ref{fig:oscillation:topo_-1V}.
		(b) The defects in \subref{fig:S_mask:topo} are masked by substituting values of pixels around the defect centers with the average value of the whole image.
		(c), (d) Fourier transforms of \qtyproduct{54 x 43}{nm} images, including \subref{fig:S_mask:topo} and \subref{fig:S_mask:topo_masked}, respectively.
		\subref{fig:S_mask:FT_masked} is the same as Fig.~\ref{fig:oscillation:FT_-1V}.
	}\label{fig:S_mask}
\end{figure}

\begin{figure}[ht]
	\centering
	\begin{subcaptiongroup}
		\includegraphics{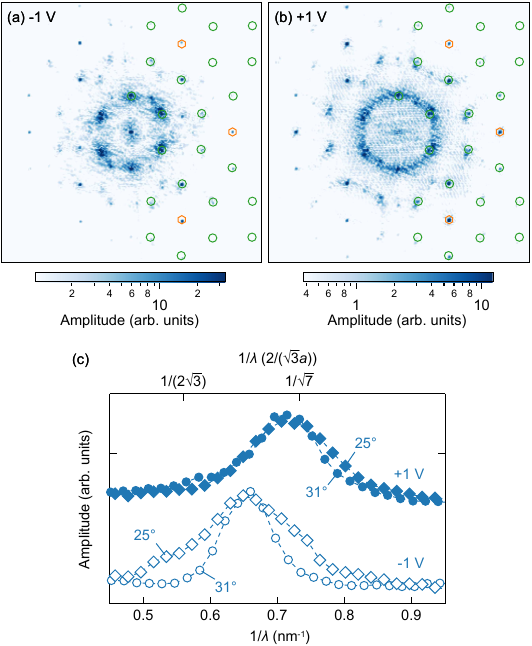}
		\phantomcaption\label{fig:S_25deg:FT_-1V}
		\phantomcaption\label{fig:S_25deg:FT_+1V}
		\phantomcaption\label{fig:S_25deg:azimuthal_average}
	\end{subcaptiongroup}
	\caption{
		The oscillation in a domain with a different angle.
		(a), (b) Fourier transforms of topographic images taken in a field of view of \qtyproduct{53 x 51}{nm}.
		The setpoint conditions are \SI{-1}{V} and \SI{0.25}{nA} for \subref{fig:S_25deg:FT_-1V} and \SI[explicit-sign=+]{1}{V} and \SI{0.5}{nA} for \subref{fig:S_25deg:FT_+1V}.
		Defects are masked before calculating \subref{fig:S_25deg:FT_-1V} as Fig.~\ref{fig:oscillation:FT_-1V}.
		The orange hexagons and the green circles depict the positions of the lattice peak of the \ce{\alpha-RuCl3} lattice and the satellite peaks, respectively.
		The satellite peak positions are calculated as described in \hyperref[sec:satellite_peaks]{Supplementary Material \ref{sec:satellite_peaks}} using $a_{\ce{\alpha-RuCl3}}=\SI{0.596}{nm}$ and $\theta=\ang{25}$.
		The markers are shown in the right half of each panel.
		(c) The azimuthal average of \subref{fig:S_25deg:FT_-1V} and \subref{fig:S_25deg:FT_+1V}.
		The pixels of the satellite peaks are removed from the average.
		Data in Fig.~\ref{fig:oscillation:azimuthal_average_incomm}, annotated as \ang{31}, are also shown for comparison.
		The peak wavenumbers of the two domains are identical for each bias voltage.
		The top axis is shown in units of the inverse of the lattice constant.
		The curves are vertically shifted for clarity.
	}\label{fig:S_25deg}
\end{figure}

\begin{figure}[ht]
	\centering
	\begin{subcaptiongroup}
		\includegraphics{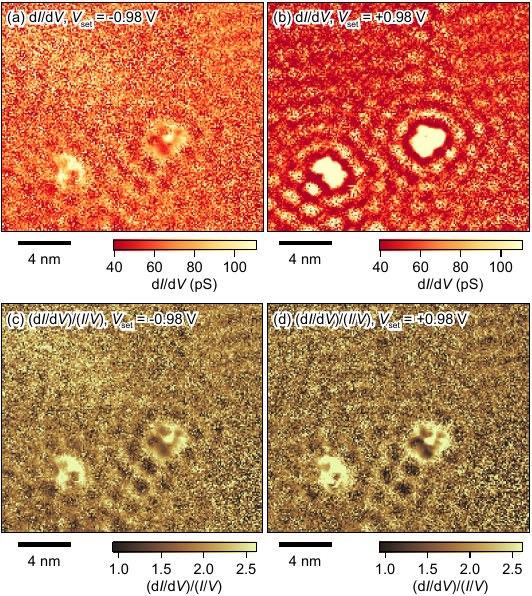}
		\phantomcaption\label{fig:S_setpoint_effect:g_negative}
		\phantomcaption\label{fig:S_setpoint_effect:g_positive}
		\phantomcaption\label{fig:S_setpoint_effect:L_negative}
		\phantomcaption\label{fig:S_setpoint_effect:L_positive}
	\end{subcaptiongroup}
	\caption{
		Mitigating the setpoint effect by normalizing $\odv{I}/{V}$.
		(a), (b) $\odv{I}/{V}$ maps at \SI{-0.98}{V} taken in the same field of view.
		The setpoint conditions are \SI{-0.98}{V} and \SI{35}{pA} for \subref{fig:S_setpoint_effect:g_negative} and \SI[explicit-sign=+]{0.98}{V} and \SI{0.1}{nA} for \subref{fig:S_setpoint_effect:g_positive}.
		Due to the setpoint effect, \subref{fig:S_setpoint_effect:g_negative} and \subref{fig:S_setpoint_effect:g_positive} are different even though the bias voltage is the same.
		(c), (d) $(\odv{I}/{V})/(I/V)$ maps at \SI{-0.98}{V} obtained by numerical division of \subref{fig:S_setpoint_effect:g_negative} and \subref{fig:S_setpoint_effect:g_positive}, respectively.
		\subref{fig:S_setpoint_effect:L_negative} and \subref{fig:S_setpoint_effect:L_positive} are almost identical, indicating that the normalization mitigates the setpoint effect.
	}\label{fig:S_setpoint_effect}
\end{figure}

\clearpage
\subsection{Statistical analysis of apparent film thickness}\label{sec:film_thickness}
We adopt statistical unsupervised learning to evaluate apparent film thickness, which offers the advantage of simultaneously performing background subtraction, estimating terrace heights, and classifying pixels, even in the presence of image deformation due to the piezo actuator~\cite{Kohsaka21}.
By utilizing all the pixels in an area to determine the apparent height of a step, this method is statistically much more robust than simply taking a line profile across the step.

\begin{figure}[ht]
	\begin{subcaptiongroup}
		\includegraphics{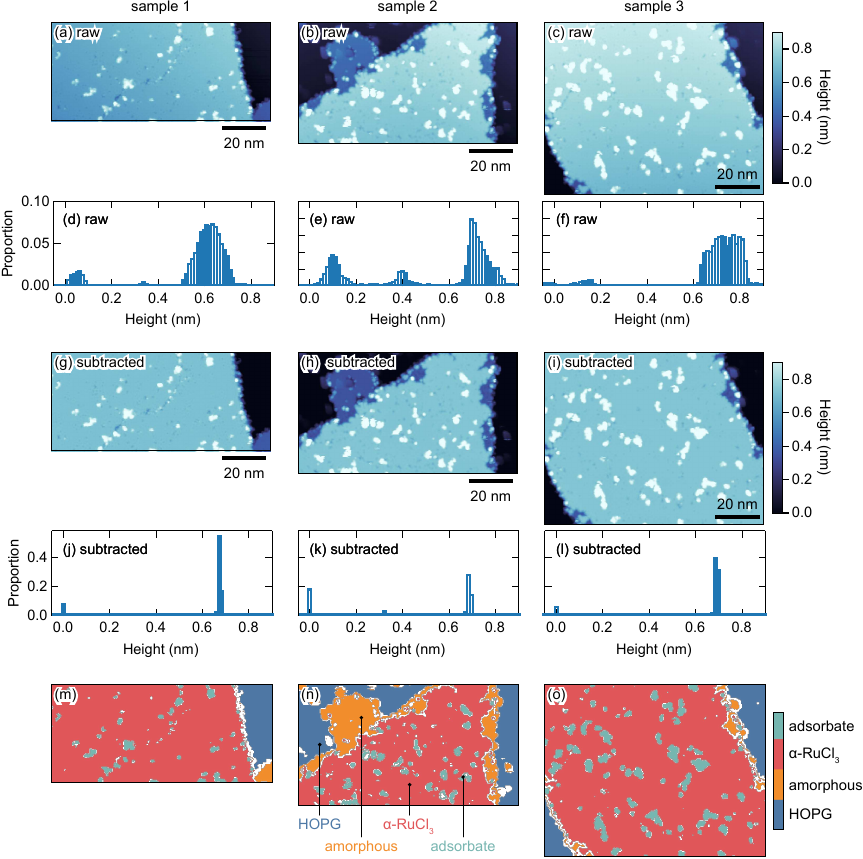}
		\phantomcaption\label{fig:S_thickness:raw1}
		\phantomcaption\label{fig:S_thickness:raw2}
		\phantomcaption\label{fig:S_thickness:raw3}
		\phantomcaption\label{fig:S_thickness:histogram_raw1}
		\phantomcaption\label{fig:S_thickness:histogram_raw2}
		\phantomcaption\label{fig:S_thickness:histogram_raw3}
		\phantomcaption\label{fig:S_thickness:subtracted1}
		\phantomcaption\label{fig:S_thickness:subtracted2}
		\phantomcaption\label{fig:S_thickness:subtracted3}
		\phantomcaption\label{fig:S_thickness:histogram_subtracted1}
		\phantomcaption\label{fig:S_thickness:histogram_subtracted2}
		\phantomcaption\label{fig:S_thickness:histogram_subtracted3}
		\phantomcaption\label{fig:S_thickness:responsibility1}
		\phantomcaption\label{fig:S_thickness:responsibility2}
		\phantomcaption\label{fig:S_thickness:responsibility3}
	\end{subcaptiongroup}
	\caption{
		(a)--(c) Raw topographic images of three different samples containing both \ce{\alpha-RuCl3} film and HOPG substrate.
		The fields of view are mostly covered by \ce{\alpha-RuCl3} films.
		HOPG substrate is shown in a dark color.
		Many adsorbates on the \ce{\alpha-RuCl3} film are shown in white.
		Some regions at the intermediate height between the HOPG substrate and the \ce{\alpha-RuCl3} film are amorphous.
		The setpoint condition was commonly \SI[explicit-sign=+]{1}{V} and \SI{20}{pA}.
		(d)--(f) Histograms of \subref{fig:S_thickness:raw1}--\subref{fig:S_thickness:raw3}, respectively.
		(g)--(i) Images subtracted background from \subref{fig:S_thickness:raw1}--\subref{fig:S_thickness:raw3}, respectively.
		(j)--(l) Histograms of \subref{fig:S_thickness:subtracted1}--\subref{fig:S_thickness:subtracted3}, respectively.
		(m)--(o) Classifications of pixels.
		The color code denotes that the responsibility for a particular class exceeds 0.8.
		Pixels with no such class are shown in white.
	}\label{fig:S_thickness}
\end{figure}

Figure~\ref{fig:S_thickness} shows the results of analysis for images taken in three different samples (Figs.~\ref{fig:S_thickness:raw1}--\ref{fig:S_thickness:raw3}).
Due to the tilting of the field of view and the deformation, the peaks in the histogram are broadened (Figs.~\ref{fig:S_thickness:histogram_raw1}--\ref{fig:S_thickness:histogram_raw3}).
By employing this statistical method, the terraces were successfully flattened and leveled, leading to much sharper peaks in the histogram (Figs.~\ref{fig:S_thickness:subtracted1}--\ref{fig:S_thickness:histogram_subtracted3}).
The success of the analysis was also confirmed by the correct classifications of pixels (Figs.~\ref{fig:S_thickness:responsibility1}--\ref{fig:S_thickness:responsibility3}).
The apparent film thicknesses are estimated at \SI{0.68}{nm}, \SI{0.69}{nm}, and \SI{0.69}{nm} for samples 1, 2, and 3, respectively.
These values are very close to the thickness in Ref.~\cite{Wang22}, corroborating that our films are monolayer \ce{\alpha-RuCl3}.

The apparent film thickness is larger than the layer height expected from the $c$-axis constant, $\SI{0.60}{nm}\times\sin(\ang{109})=\SI{0.57}{nm}$~\cite{Johnson15,Cao16}.
Note that the metallic density of states of the substrate \ac{HOPG} is not the reason for the larger apparent thickness; it results in a smaller apparent thickness.
The first layer adjacent to the substrate is also observed to be thicker than other layers in exfoliated samples~\cite{Du18}.
Since a thicker first layer is observed in samples obtained by three different methods, pulsed laser deposition, molecular beam epitaxy, and exfoliation, it likely indicates weak coupling between the film and the substrate.

\subsection{Local amplitude map of the decaying oscillation}\label{sec:local_amplitude_map}
When the amplitude and phase of a modulation vary gradually in an image $f(\bm{r})$, their spatial variations can be extracted as the amplitude and phase of $\mathcal{L}\ab[f(\bm{r})e^{i\bm{q}\cdot\bm{r}}]$, where $\mathcal{L}$ is a low-pass filter, and $\bm{q}$ is the wavevector of the modulation.

Define $a_{\bm{q}}(\bm{r})$ and $\theta_{\bm{q}}(\bm{r})$ as the spatial variation of the amplitude and the phase, respectively, that we aim to extract.
The spatial variation of the modulation arises from wavepackets centered around $\bm{q}$ and $-\bm{q}$,
\begin{align}
	\operatorname{Re}\ab\{2a_{\bm{q}}(\bm{r})e^{i\ab[\bm{q}\cdot\bm{r}+\theta_{\bm{q}}(\bm{r})]}\}
	&= \sum_{\ab|\bm{k}-\bm{q}|<k_0}F(\bm{k})e^{i\bm{k}\cdot\bm{r}}
	 + \sum_{\ab|\bm{k}+\bm{q}|<k_0}F(\bm{k})e^{i\bm{k}\cdot\bm{r}}
	 \label{eq:LAM:varying_modulation}
\end{align}
where $k_0$ is a small wavenumber satisfying $k_0 < \ab|\bm{q}|$, and $F(\bm{k})$ is the Fourier transform of $f(\bm{r})$.
The image $f(\bm{r})$ is the sum of the modulation and other components,
\begin{align}
	f(\bm{r}) = \operatorname{Re}\ab\{2a_{\bm{q}}(\bm{r})e^{i\ab[\bm{q}\cdot\bm{r}+\theta_{\bm{q}}(\bm{r})]}\}
		+ \sum_{\substack{\ab|\bm{k}-\bm{q}|\geq k_0\\\ab|\bm{k}+\bm{q}|\geq k_0}}
			F(\bm{k})e^{i\bm{k}\cdot\bm{r}}.
\end{align}
To extract $a_{\bm{q}}(\bm{r})$ and $\theta_{\bm{q}}(\bm{r})$ from $f(\bm{r})$, we initially multiply $e^{-i\bm{q}\cdot\bm{r}}$ to $f(\bm{r})$
\begin{align}
	f(\bm{r})e^{-i\bm{q}\cdot\bm{r}}
	&= \sum_{\ab|\bm{k}|<k_0} F(\bm{k}+\bm{q})e^{i\bm{k}\cdot\bm{r}}
	 + \sum_{\ab|\bm{k}+2\bm{q}|<k_0} F(\bm{k}+\bm{q})e^{i\bm{k}\cdot\bm{r}}
	 + \sum_{\substack{\ab|\bm{k}|\geq k_0\\\ab|\bm{k}+2\bm{q}|\geq k_0}}
		F(\bm{k}+\bm{q})e^{i\ab\bm{k}\cdot\bm{r}}.
\end{align}
Subsequently, applying a low-pass filter $\mathcal{L}$ retains only the first term,
\begin{align}
	\mathcal{L}\ab[f(\bm{r})e^{-i\bm{q}\cdot\bm{r}}]
	= \sum_{\ab|\bm{k}|< k_0}F(\bm{k}+\bm{q})e^{i\bm{k}\cdot\bm{r}}.
\end{align}
Since $f(\bm{r})$ is real, $F(\bm{k})=F^*(-\bm{k})$, which leads to Eq.~\eqref{eq:LAM:varying_modulation} becoming
\begin{align}
	\operatorname{Re}\ab\{2a_{\bm{q}}(\bm{r})e^{i\ab[\bm{q}\cdot\bm{r}+\theta_{\bm{q}}(\bm{r})]}\}
	&= a_{\bm{q}}(\bm{r})\ab\{
		e^{i\ab[\bm{q}\cdot\bm{r}+\theta_{\bm{q}}(\bm{r})]}
	  + e^{-i\ab[\bm{q}\cdot\bm{r}+\theta_{\bm{q}}(\bm{r})]}
	\} \\
	&= \sum_{\ab|\bm{k}|<k_0}\ab[
		F(\bm{k}+\bm{q})e^{i\ab(\bm{k}+\bm{q})\cdot\bm{r}}
		+ F^*(\bm{k}+\bm{q})e^{-i\ab(\bm{k}+\bm{q})\cdot\bm{r}}
		].
\end{align}
Therefore,
\begin{align}
	a_{\bm{q}}(\bm{r})e^{i\theta_{\bm{q}}(\bm{r})}
	= \sum_{\ab|\bm{k}|<k_0} F(\bm{k}+\bm{q})e^{i\bm{k}\cdot\bm{r}}
	= \mathcal{L}\ab[f(\bm{r})e^{-i\bm{q}\cdot\bm{r}}].
\end{align}
The amplitude and the phase to be extracted correspond to those of $\mathcal{L}\ab[f(\bm{r})e^{-i\bm{q}\cdot\bm{r}}]$.
Although we utilize only the amplitude map for the current analysis, the phase map can be used to correct image distortions when selecting wavevectors of a known lattice~\cite{Lawler10}.
In this context, the above method is a generalization of the so-called Lawler-Fujita algorithm.

\begin{figure}[ht]
	\begin{subcaptiongroup}
		\includegraphics{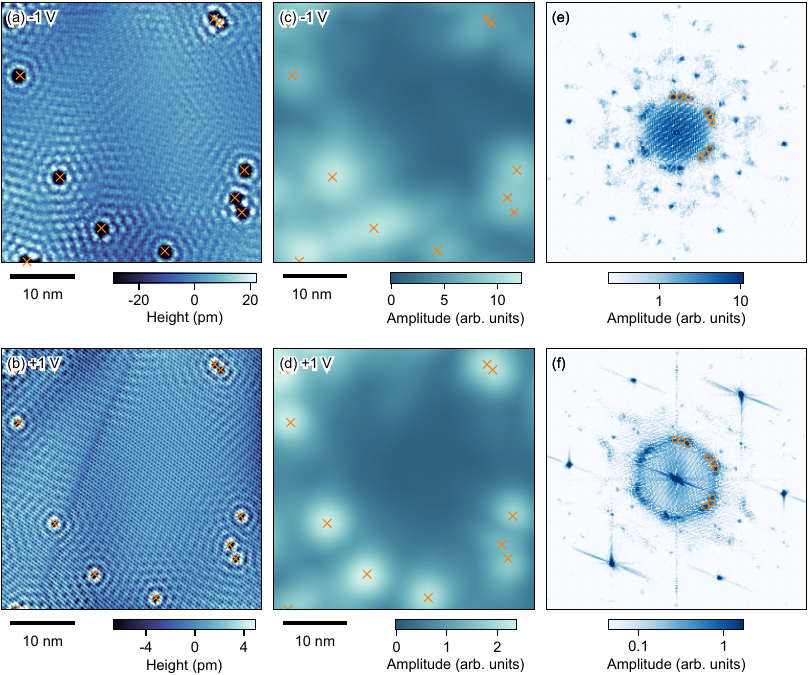}
		\phantomcaption\label{fig:S_LAM:topo_-1V}
		\phantomcaption\label{fig:S_LAM:topo_+1V}
		\phantomcaption\label{fig:S_LAM:LAM_-1V}
		\phantomcaption\label{fig:S_LAM:LAM_+1V}
		\phantomcaption\label{fig:S_LAM:FT_-1V}
		\phantomcaption\label{fig:S_LAM:FT_+1V}
	\end{subcaptiongroup}
	\caption{
		(a), (b) Topographic images shown in Figs.~\ref{fig:oscillation:topo_-1V} and \ref{fig:oscillation:topo_+1V}, respectively, with the orange markers indicating the positions of defect centers.
		(c), (d) The local amplitude maps same as Figs.~\ref{fig:oscillation:LAM_-1V} and \ref{fig:oscillation:LAM_+1V}, respectively.
		The orange markers are placed at the same locations as those in \subref{fig:S_LAM:topo_-1V} and \subref{fig:S_LAM:topo_+1V}.
		(e), (f) The Fourier transforms of areas including \subref{fig:S_LAM:topo_-1V} and \subref{fig:S_LAM:topo_+1V}, respectively.
		The center and radius of the orange circles indicate the wavevector of the modulation and the half-width at half-maximum of the Gaussian low-pass filter, respectively.
	}\label{fig:S_local_amplitude_map}
\end{figure}

Figure~\ref{fig:S_local_amplitude_map} shows the local amplitude maps of the decaying oscillation for the topographic images shown in Figs.~\ref{fig:oscillation:topo_-1V} and \ref{fig:oscillation:topo_+1V}.
Figs.~\ref{fig:S_LAM:LAM_-1V} and \ref{fig:S_LAM:LAM_+1V} are the sum of local ampiltude maps for multiple wavevectors, that is $\sum_{\bm{q}}a_{\bm{q}}(\bm{r})$.
The centers of the orange circles in Figs.~\ref{fig:S_LAM:FT_-1V} and \ref{fig:S_LAM:FT_+1V} indicate the positions of the wavevectors.
As the low-pass filter, we employ a Gaussian filter.
The radius of the orange circles (\SI{0.06}{nm^{-1}}) depicts the half-width at half-maximum of the Gaussian filter.

\subsection{Excluding commensurate structures}\label{sec:possible_commensurate_structure}

\begin{figure}[ht]
	\centering
	\begin{subcaptiongroup}
		\includegraphics{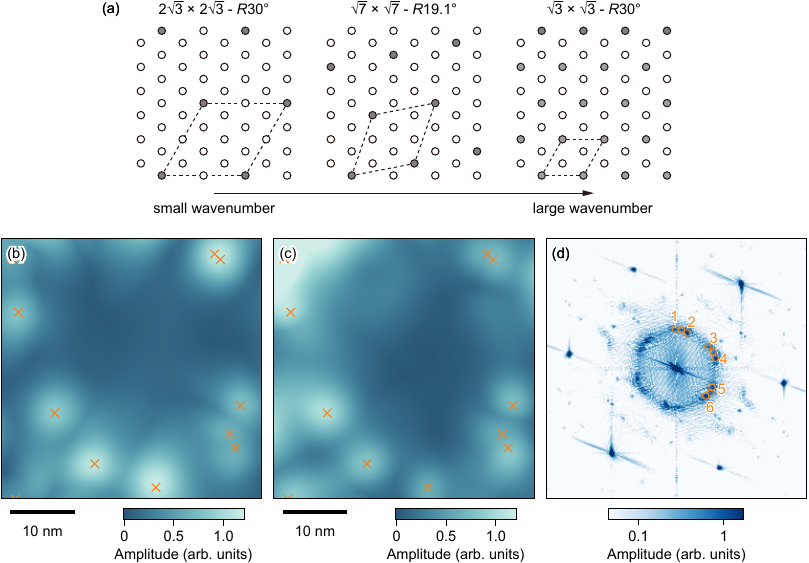}
		\phantomcaption\label{fig:S_sqrt7:superlattices}
		\phantomcaption\label{fig:S_sqrt7:135}
		\phantomcaption\label{fig:S_sqrt7:246}
		\phantomcaption\label{fig:S_sqrt7:FT}
	\end{subcaptiongroup}
	\caption{
		(a) Commensurate superlattice structures with the wavevectors near that of the oscillation.
		(b), (c) The local amplitude maps calculated for the sets of peaks $\{1, 3, 5\}$ for \subref{fig:S_sqrt7:135} and $\{2, 4, 6\}$ for \subref{fig:S_sqrt7:246}, respectively.
		The peak numbers are indicated in \subref{fig:S_sqrt7:FT}.
		(d) The Fourier transform same as shown in Fig.~\ref{fig:oscillation:FT_+1V}.
		The numbers next to the orange circles denote peaks used for calculating \subref{fig:S_sqrt7:135} and \subref{fig:S_sqrt7:246}.
	}
	\label{fig:S_sqrt7}
\end{figure}

There are three commensurate superlattice structures with the wavevectors near that of the oscillation: in order of increasing wavenumber, $(2\sqrt{3}\times 2\sqrt{3})-R\ang{30}$, $(\sqrt{7}\times\sqrt{7})-R\ang{19.1}$, and $(\sqrt{3}\times\sqrt{3})-R\ang{30}$, as shown in Fig.~\ref{fig:S_sqrt7:superlattices}.
The comparison of the wavenumbers shown in Fig.~\ref{fig:oscillation:azimuthal_average_incomm} implies a possibility that the oscillation at \SI[explicit-sign=+]{1}{V} may have the $(\sqrt{7}\times\sqrt{7})-R\ang{19.1}$ structure.

If such a commensurate structure were present, Fig.~\ref{fig:oscillation:FT_+1V} would suggest that oscillations around certain defects exhibit only the $(\sqrt{7}\times\sqrt{7})-R\ang[explicit-sign=+]{19.1}$ structure, while others display solely the $\ang{-19.1}$ counterpart.
A pattern around a defect containing both $\ang{\pm 19.1}$ components is incommensurate.
To investigate this possibility, we analyzed the local amplitude map separately for the sets of possible \ang[explicit-sign=+]{19.1} and \ang{-19.1} peaks, as illustrated in Figs.~\ref{fig:S_sqrt7:135}--\ref{fig:S_sqrt7:FT}.
The analysis shows that the oscillations around all defects include contributions from both $\ang{\pm 19.1}$ components, demonstrating that the $(\sqrt{7}\times\sqrt{7})-R\ang{19.1}$ commensurate structure is not present.

\subsection{How to calculate positions of satellite peaks}\label{sec:satellite_peaks}
First, we describe how to calculate the Fourier transforms as shown in Figs.~\ref{fig:oscillation:FT_-1V} and \ref{fig:oscillation:FT_+1V}.
Let $\bm{G}^{\mathrm{obs}(j)}_{\ce{\alpha-RuCl3}}$ $(j = 1,\ 2,\ 3)$ be positions of fundamental lattice peaks observed in a Fourier transform of a topographic image of an \ce{\alpha-RuCl3} film.
Generally, due to image deformation caused by a piezo actuator to scan a tip, $\ab|\bm{G}^{\mathrm{obs}(j)}_{\ce{\alpha-RuCl3}}|\neq\ab|\bm{G}^{\mathrm{obs}(k)}_{\ce{\alpha-RuCl3}}|$ and the angle between $\bm{G}^{\mathrm{obs}(j)}_{\ce{\alpha-RuCl3}}$ and $\bm{G}^{\mathrm{obs}(k)}_{\ce{\alpha-RuCl3}}$ is neither \ang{60} nor \ang{120} even if they are the case.
Nevertheless, when we calculate a Fourier transform of an image, we assume that the \ce{\alpha-RuCl3} lattice maintains $C_3$ symmetry and apply a linear transformation such that lattice peaks reside at the ideal locations,
\begin{align}
	\bm{G}^{\mathrm{ideal}(1)}_{\ce{\alpha-RuCl3}} =
		\begin{pmatrix}
			\dfrac{1}{\sqrt{3}a_{\ce{\alpha-RuCl3}}} \\[2ex]
			\dfrac{1}{a_{\ce{\alpha-RuCl3}}}
		\end{pmatrix},\quad
	\bm{G}^{\mathrm{ideal}(2)}_\mathrm{\ce{\alpha-RuCl3}} =
		\begin{pmatrix}
			\dfrac{2}{\sqrt{3}a_{\ce{\alpha-RuCl3}}} \\[2ex]
			0
		\end{pmatrix},\quad
	\bm{G}^{\mathrm{ideal}(3)}_{\ce{\alpha-RuCl3}} =
		\begin{pmatrix}
			\dfrac{1}{\sqrt{3}a_{\ce{\alpha-RuCl3}}} \\[2ex]
			-\dfrac{1}{a_{\ce{\alpha-RuCl3}}}
		\end{pmatrix},
\end{align}
where $a_{\ce{\alpha-RuCl3}}$ is the $a$-axis lattice constant of \ce{\alpha-RuCl3}.
(The validity of this transformation will be assessed subsequently.)
The transformation matrix $M$ is given by the peak positions,
\begin{align}
	A_\mathrm{ideal} &= MA_\mathrm{obs}, \\
	A_\mathrm{ideal} &= \begin{pmatrix}
		\bm{G}^{\mathrm{ideal}(1)}_{\ce{\alpha-RuCl3}} &
		\bm{G}^{\mathrm{ideal}(2)}_{\ce{\alpha-RuCl3}} &
		\bm{G}^{\mathrm{ideal}(3)}_{\ce{\alpha-RuCl3}}
	\end{pmatrix},\\
	A_\mathrm{obs} &= \begin{pmatrix}
		\bm{G}^{\mathrm{obs}(1)}_{\ce{\alpha-RuCl3}} &
		\bm{G}^{\mathrm{obs}(2)}_{\ce{\alpha-RuCl3}} &
		\bm{G}^{\mathrm{obs}(3)}_{\ce{\alpha-RuCl3}}
	\end{pmatrix}.
\end{align}
$M$ is determined in a least-squares manner,
\begin{align}
	M = {A_\mathrm{ideal}}{A_\mathrm{obs}^\top}(A_\mathrm{obs}A_\mathrm{obs}^\top)^{-1}.
\end{align}
A wavevector after the transformation $\bm{q}^\prime$ is related to one before the transformation $\bm{q}$ as
\begin{align}
	\bm{q}^\prime = M\bm{q}.
\end{align}
Let $F^\prime$ and $F$ be Fourier transform after and before the linear transformation.
Since $F^\prime(\bm{q}^\prime) = F(\bm{q})$,
\begin{align}
	F^\prime(\bm{q}^\prime) = \sum_{\bm{r}} f(\bm{r})\exp\ab(i\bm{q}\cdot\bm{r}) = \sum_{\bm{r}} f(\bm{r})\exp\ab(iM^{-1}\bm{q}^\prime\cdot\bm{r}),
\end{align}
where $f(\bm{r})$ is an original image in real space.

\begin{figure}[ht]
	\centering
	\begin{subcaptiongroup}
		\includegraphics{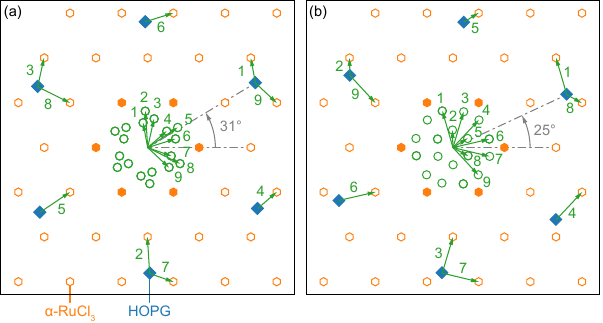}
		\phantomcaption\label{fig:S_green_peak:31}
		\phantomcaption\label{fig:S_green_peak:25}
	\end{subcaptiongroup}
	\caption{
		The relationship between the positions of satellite peaks near the origin and the lattice peak positions.
		$\theta=\ang{31}$ for \subref{fig:S_green_peak:31} and \ang{25} for \subref{fig:S_green_peak:31}.
		The orange hexagons and blue squares denote the peak positions of \ce{\alpha-RuCl3} and \ac{HOPG} lattices, respectively.
		The fundamental peaks are shown in filled markers, while the harmonic peaks are shown in open markers.
		The green circles indicate the calculated positions of satellite peaks near the origin.
		The differences from the HOPG peaks to the nearest \ce{\alpha-RuCl3} peaks correspond to the peak positions of satellite peaks near the origin.
	}\label{fig:S_green_peak}
\end{figure}

Next, we try to assign the peaks marked by the green circles in Figs.~\ref{fig:oscillation:FT_-1V} and \ref{fig:oscillation:FT_+1V}.
Henceforth, we refer to the peaks as the green peaks for simplicity.
Suppose the green peaks are satellite peaks between the \ce{\alpha-RuCl3} and \ac{HOPG} lattices.
In that case, their positions are given by a linear combination of the peak positions of the two lattices.
Since harmonic peaks of the \ce{\alpha-RuCl3} lattice are observed in the Fourier transform, the position of a satellite peak $\bm{q}_\mathrm{satellite}$ is
\begin{align}
	\bm{q}_\mathrm{satellite}
	= \sum_j m_j\bm{G}^{\mathrm{ideal}(j)}_\mathrm{\ce{\alpha-RuCl3}}
	\pm \sum_k R(\theta)\bm{G}^{\mathrm{ideal}(k)}_\mathrm{HOPG},
\end{align}
where $m_j$ is an integer.
$\bm{G}^{\mathrm{ideal}(k)}_\mathrm{HOPG}$ $(k = 1,\ 2,\ 3)$ are positions of fundamental peaks of the \ac{HOPG} lattice at the ideal locations,
\begin{align}
	\bm{G}^{\mathrm{ideal}(1)}_\mathrm{HOPG} =
		\begin{pmatrix}
			\dfrac{1}{\sqrt{3}a_\mathrm{HOPG}} \\[2ex]
			\dfrac{1}{a_\mathrm{HOPG}}
		\end{pmatrix},\quad
	\bm{G}^{\mathrm{ideal}(2)}_\mathrm{HOPG} =
		\begin{pmatrix}
			\dfrac{2}{\sqrt{3}a_\mathrm{HOPG}} \\[2ex]
			0
		\end{pmatrix},\quad
	\bm{G}^{\mathrm{ideal}(3)}_\mathrm{HOPG} =
		\begin{pmatrix}
			\dfrac{1}{\sqrt{3}a_\mathrm{HOPG}} \\[2ex]
			-\dfrac{1}{a_\mathrm{HOPG}}
		\end{pmatrix},
\end{align}
where $a_\mathrm{HOPG}=\SI{0.246}{nm}$ is the $a$-axis lattice constant of \ac{HOPG}~\cite{Baskin55}.
$\theta$ is the angle of \ac{HOPG} lattice relative to the \ce{\alpha-RuCl3} lattice, and $R(\theta)$ is a rotation matrix,
\begin{align}
	R(\theta) = \begin{pmatrix}
		\cos\ab(\theta) & -\sin\ab(\theta) \\
		\sin\ab(\theta) & \cos\ab(\theta)
	\end{pmatrix}.
\end{align}
The positions of green markers in Figs.~\ref{fig:oscillation:FT_-1V} and \ref{fig:oscillation:FT_+1V} were obtained for various combinations of $m_1$, $m_2$, and $m_3$ using and $\theta=\ang{31}$ and $a_{\ce{\alpha-RuCl3}}=\SI{0.596}{nm}$~\cite{Banerjee16}.
The calculated peak positions are also close to the observed satellite peaks for $a_{\ce{\alpha-RuCl3}}=\SI{0.597}{nm}$ or \SI{0.598}{nm}~\cite{Johnson15,Cao16}.
Fig.~\ref{fig:S_green_peak} shows the relationship between the positions of green peaks near the origin and the lattice peak positions.
Since the positions of green peaks are calculated from harmonic peaks of the \ce{\alpha-RuCl3} lattice and the fundamental peaks of the \ac{HOPG} lattice, they are sensitive to $\theta$.
Given that this simple approach well explains the positions of green peaks, it is reasonable to affirm that the \ce{\alpha-RuCl3} lattice retains $C_3$ symmetry and that the green peaks are satellite peaks between \ce{\alpha-RuCl3} and \ac{HOPG} lattices.
However, there remains uncertainty for $a_{\ce{\alpha-RuCl3}}$ and $\theta$ as described in \hyperref[sec:lattice_constant]{Supplementary Material \ref{sec:lattice_constant}}.

\subsection{The lattice constant of \ce{\alpha-RuCl3} films}\label{sec:lattice_constant}
As described at the beginning of \hyperref[sec:satellite_peaks]{Supplementary Material \ref{sec:satellite_peaks}}, images are always deformed due to the piezo actuator.
Since the deformation depends on the history of tip positions (such as scan speed, scan direction, and where previous images were taken), it varies from one image to another and can not be determined a priori.
Therefore, unless a reference lattice is imaged simultaneously, it is impossible to accurately determine the absolute value of the lattice constant of \ce{\alpha-RuCl3} films even if the piezo actuator was calibrated in advance.

\begin{figure}[ht]
	\centering
	\includegraphics[scale=1.5]{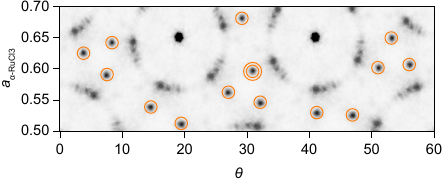}
	\caption{
		The intensity of the Fourier transform at satellite peaks calculated for various $a_{\ce{\alpha-RuCl3}}$ and $\theta$.
		The original real-space image is the same as Fig.~\ref{fig:oscillation:FT_-1V}.
		The positions of satellite peaks are well explained for the combinations marked by the orange circles.
		The double circle indicates $a_{\ce{\alpha-RuCl3}}=\SI{0.596}{nm}$ and $\theta=\ang{31}$, which is used for the in-place calibration in the main text.
	}\label{fig:S_lattice_constant}
\end{figure}

In our case, although the satellite peaks are observed, the fundamental lattice peaks of \ac{HOPG} are not observed.
Because of the lack of a reference lattice, we can not exclusively conclude that the $a$-axis constant of the \ce{\alpha-RuCl3} film is the same as the bulk.
Figure~\ref{fig:S_lattice_constant} shows the intensity of the Fourier transform at the positions of satellite peaks calculated for various $a_{\ce{\alpha-RuCL3}}$ and $\theta$.
The satellite peak positions are explained for many combinations of $a_{\ce{\alpha-RuCl3}}$ and $\theta$, marked by orange circles.
In the main text, we use $a_{\ce{\alpha-RuCl3}}=\SI{0.596}{nm}$ and $\theta=\ang{31}$ (indicated by the double orange circle in Fig.~\ref{fig:S_lattice_constant}) for the in-place calibration for simplicity.
In any case, however, $\theta$ is neither \ang{0} nor \ang{30}, as indicated by the positions of satellite peaks away from the high symmetry directions.

\subsection{Majorana-based scenario for the incommensurate spatial oscillation}\label{sec:Majorana_scenario}

\begin{figure}[ht]
	\centering
	\begin{subcaptiongroup}
		\includegraphics{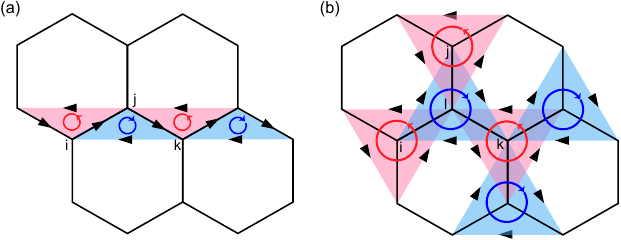}
		\phantomcaption\label{fig:S_orbital_current:current1}
		\phantomcaption\label{fig:S_orbital_current:current2}
	\end{subcaptiongroup}
	\caption{
		Examples of circular orbital currents which generate the scalar spin chirality.
		The orbital currents flow in staggered directions to preserve zero total angular momentum.
		Here, staggered patterns compatible with the sub-lattice structure of the honeycomb lattice are shown.
		Note that staggered patterns which are not compatible with the sub-lattice structure do not affect low-energy features of the Majorana band qualitatively.
	}\label{fig:S_orbital_current}
\end{figure}

\begin{figure}[ht]
	\centering
	\begin{subcaptiongroup}
		\includegraphics{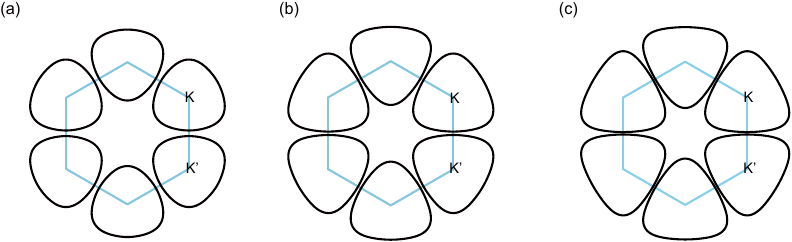}
		\phantomcaption\label{fig:S_Majorana_Fermi_surface:1.5}
		\phantomcaption\label{fig:S_Majorana_Fermi_surface:2.5}
		\phantomcaption\label{fig:S_Majorana_Fermi_surface:3.0}
	\end{subcaptiongroup}
	\caption{
		The Majorana Fermi surfaces of the model Hamiltonian $\mathcal{H}$ in the case without vacancies.
		$\chi_0 = 1.5$ for \subref{fig:S_Majorana_Fermi_surface:1.5}, 2.5 for \subref{fig:S_Majorana_Fermi_surface:2.5}, and 3.0 for \subref{fig:S_Majorana_Fermi_surface:3.0}.
		The energy unit is $K=1$ for all numerical calculations in Supplementary material.
		In these calculations, a non-Kitaev interaction term $\mathcal{H}_\mathrm{NK}$ is neglected.
	}\label{fig:S_Majorana_Fermi_surface}
\end{figure}

\begin{figure}[ht]
	\centering
	\begin{subcaptiongroup}
		\includegraphics{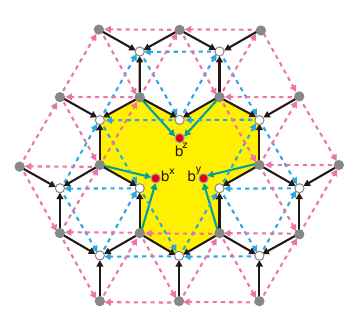}
	\end{subcaptiongroup}
	\caption{
		Hopping processes of itinerant Majorana fermions.
		The additional gauge Majorana fields $b^x$, $b^y$, and $b^z$ appear at the sites neighboring the vacancy site.
		A vison trapped at the vacancy is denoted in yellow color.
	}\label{fig:S_Majorana_Fermion_Hopping}
\end{figure}

In this section, we present a scenario based on Majorana physics for the origin of the incommensurate oscillation observed via the \ac{STM} measurements.
Although the bias voltage imposed on the \ac{STM} tip is larger than the Mott gap, electrons carrying the tunnel current injected from the \ac{STM} tip may generally decay into low-energy excitations in the Kitaev monolayer, and the local density of states probed via the \ac{STM} measurements is strongly affected by the spatial distributions of these excitations, showing the trace of the spin liquid state.
There are two types of excitations in the Kitaev \ac{QSL}; itinerant Majorana fermions and visons~\cite{Kitaev06}.
The incommensurate spatial structure of the density of states implies that excitations associated with this oscillation have a characteristic length scale akin to a Fermi wavelength.
Then, it is natural to expect that the itinerant Majorana fermions play an essential role in the incommensurate oscillation, because of their fermionic character.
However, for the pure Kitaev model, the Fermi wavelength of the itinerant Majorana fermions is commensurate, because the Dirac points of the Majorana band cross zero-energy at K and K$^\prime$ points in the Brillouin zone.
Nevertheless, it is possible to realize the Majorana Fermi surface with incommensurate Fermi wavenumbers, if there are perturbations which break symmetries protecting the positions of the Dirac points.
According to the topological argument based on the twisted $K$ theory, in the case of the Kitaev model, these symmetries are time-reversal symmetry and inversion symmetry~\cite{Shiozaki22,Takikawa19}.
With this insight, we postulate that tunnel currents from the \ac{STM} tip, which break both time-reversal and inversion symmetries, induce circular orbital currents in the \ce{RuCl3} monolayer, which flow in staggered directions on the honeycomb lattice since the total orbital angular momentum on the monolayer plane must be kept zero.
(See Figs.~\ref{fig:S_orbital_current:current1} and \ref{fig:S_orbital_current:current2}.)
The orbital currents flowing on edges of a triangular in the honeycomb lattice induce scalar spin chirality $\chi_{ijk}\equiv \bm{S}_i\cdot(\bm{S}_j\times\bm{S}_k)$ because of the inverse effect of the circular current generation due to spin chirality in a Mott insulator, $J\propto \chi_{ijk}$~\cite{Bulaevskii08}.
The perturbation term of the Hamiltonian due to this effect is given by $\mathcal{H}_{\rm SSC}\propto \sum_{i,j,k}\chi_{ijk}$.
For the staggered orbital currents shown in Fig.~\ref{fig:S_orbital_current:current1}, the associated spin chirality yields the next nearest-neighbor hopping term of itinerant Majorana fermions, since $ \chi_{ijk} \sim u^{\alpha}_{ij}u^{\beta}_{kj}ic_kc_i$ in the Majorana representation of the Kitaev \ac{QSL} state.
Here, $c_j$ is the operator of an itinerant Majorana fermion, $u_{jk}^{\alpha}$ ($\alpha=x,y,z$) is a $Z_2$ gauge field, and $i$, $k$ are
next-nearest neighbor sites.
On the other hand, for the current configuration shown in Fig.~\ref{fig:S_orbital_current:current2}, the spin chirality results in four-Majorana interaction terms, \textit{e.g.} $ \chi_{ijk} \sim u_{il}^xu_{kl}^yu_{jl}^zc_ic_jc_kc_l$.
In the mean-field approximation, these terms also give rise to next-nearest neighbor hopping terms, \textit{e.g.} $\sim u_{il}^xu_{kl}^yu_{jl}^zc_kc_i$.
Notably, because of the alternating sign of the next nearest-neighbor hopping terms, these perturbations shift the position of the Dirac points of the Majorana band away from zero energy, resulting in the formation of Majorana Fermi surfaces with incommensurate Fermi wavenumbers.
Then, the incommensurate spatial oscillation of the density of states similar to the Friedel oscillation can occur, when itinerant Majorana fermions are scattered at vacancy sites.
We emphasize that although the perturbation due to the \ac{STM} current is local, and the induced spin chirality decays exponentially, the incommensurate Fermi wavenumbers can be generated in the vicinity of the \ac{STM} tip, if the decay length of the spin chirality is just a few times larger than the lattice constant.
To demonstrate this scenario, we performed model calculations with the Majorana Hamiltonian derived from the generalized Kitaev model, $\mathcal{H}=\mathcal{H}_\mathrm{K}+\mathcal{H}_\mathrm{SSC}+\mathcal{H}_\mathrm{NK}$, where $\mathcal{H}_\mathrm{K}$ is the Hamiltonian of the pure Kitaev model~\cite{Kitaev06}, and $\mathcal{H}_\mathrm{NK}$ is a non-Kitaev interaction term.
The expressions of $\mathcal{H}_\mathrm{K}$ and $\mathcal{H}_\mathrm{SSC}$ are given by,
\begin{gather}
	\mathcal{H}_\mathrm{K} = K\sum_{\braket[1]{jk}_{\alpha}}u^{\alpha}_{jk}ic_jc_k, \\[8pt]
	\mathcal{H}_\mathrm{SSC} = \chi_0\sum_{\langle\!\langle jk\rangle\!\rangle}(-1)^l\mathcal{T}(\bm{r}_{ij}-\bm{r}_\mathrm{STM})u^{\alpha}_{lj}u^{\beta}_{lk}ic_jc_k,
	\label{eq:hssc}
\end{gather}
where $K$ is the Kitaev interaction strength which defines the energy unit in our numerical calculations, and $\chi_0$ is a coupling constant.
In the expression of $\mathcal{H}_\mathrm{SSC}$, the sign factor $(-1)^l$ arises from the staggered configuration of scalar spin chirality.
This feature is crucially important for the realization of Majorana Fermi surfaces.
$\mathcal{T}(\bm{r}_{ij}-\bm{r}_\mathrm{STM})$ is a damping factor which describes the decay of the spin chirality as a function of the distance between the position, $\bm{r}_{ij}=(\bm{r}_i+\bm{r}_j)/2$, and that of the \ac{STM} tip, $\bm{r}_\mathrm{STM}$.
Since the spin chirality is induced by the tunnel currents from the \ac{STM} tip in this scenario, it should decay as the distance from the \ac{STM} tip increases.
This effect is incorporated in the damping factor.
We use the gaussian form $\mathcal{T}(\bm{r})=e^{-\frac{|\bm{r}|^2}{\ell_0^2}}$ for numerical calculations.
We also assume that $\mathcal{H}_\mathrm{NK}$ is sufficiently small, and does not affect the stability of the spin liquid state.
However, we should note that in the case of a vacancy, $\mathcal{H}_\mathrm{NK}$ gives an important effect on a bound Majorana state in the vicinity of a vacancy, which will be discussed later.
As mentioned above, $\mathcal{H}_\mathrm{SSC}$ shifts the position of Dirac points of the itinerant Majorana band away from zero energy, which realizes the Majorana Fermi surfaces with incommensurate Fermi wavenumbers, as shown in Figs.~\ref{fig:S_Majorana_Fermi_surface:1.5}, \ref{fig:S_Majorana_Fermi_surface:2.5}, and \ref{fig:S_Majorana_Fermi_surface:3.0}.
We, now, examine the effects of a lattice vacancy which plays the role of a scattering potential acting on the itinerant Majorana fermions.
An important observation is that at the nearest neighbor sites of a vacancy in the Kitaev model, isolated gauge Majorana fields $b^{\alpha}_j$, which do not couple to any other Majorana particles in the Kitaev \ac{QSL}, appear, and furthermore, these gauge Majorana fields are coupled to nearest-neighbor itinerant Majorana fields via $\mathcal{H}_\mathrm{NK}$.
(See Fig.~\ref{fig:S_Majorana_Fermion_Hopping}.)
Thus, even when $\mathcal{H}_\mathrm{NK}$ is small enough, the gauge Majorana fields near the vacancy affect crucially scattering processes of itinerant Majorana fermions due to the vacancy.
To take this effect into account, we assume the form of $\mathcal{H}_\mathrm{NK}$ as,
\begin{eqnarray}
	\mathcal{H}_\mathrm{NK}=t_{bc}\sum_{\braket[1]{jk}_{\alpha}}u^{\alpha}_{jk}ib^{\gamma}_jc_k,
	\label{eq:hnk}
\end{eqnarray}
which describes a minimal coupling between the gauge Majoranas and itinerant Majoranas.
Here, the superscript $\gamma$ of the gauge Majorana field $b^{\gamma}_j$ means that the vacancy site is connected to the site $j$ via the $\gamma$-bond ($\gamma=x,y,z$).
Note that Eq.~(\ref{eq:hnk}) is microscopically derived from the 1st order perturbative expansion with respect to the symmetric off-diagonal exchange interaction, the $\Gamma^\prime$ term.
It is noted that the 1st order corrections due to the $\Gamma^\prime$ term do not affect the bulk spin liquid state away from the vacancy site.

\begin{figure}[ht]
	\centering
	\begin{subcaptiongroup}
		\includegraphics{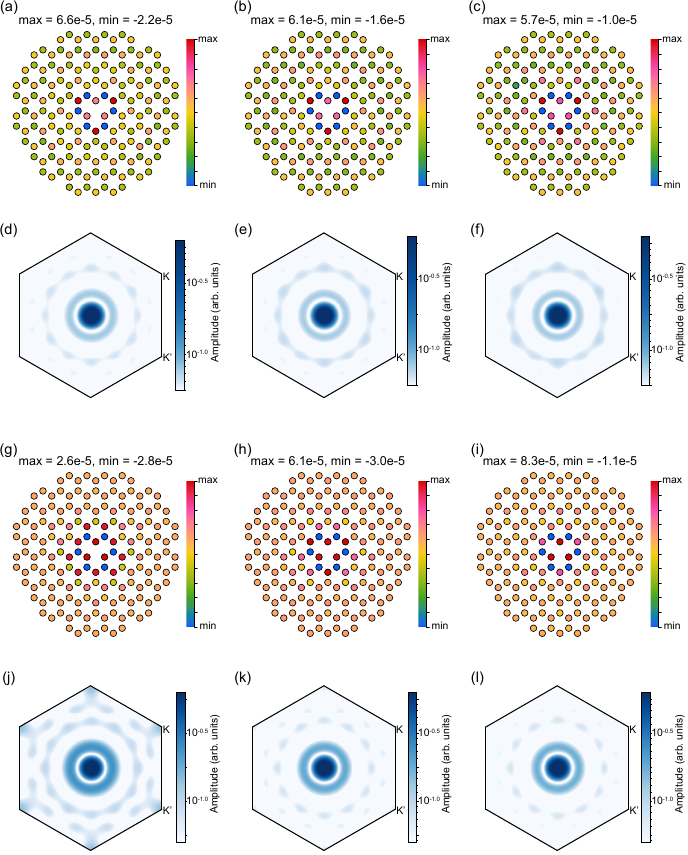}
		\phantomcaption\label{fig:S_charge_density:bound_2.5_real}
		\phantomcaption\label{fig:S_charge_density:bound_3.0_real}
		\phantomcaption\label{fig:S_charge_density:bound_3.5_real}
		\phantomcaption\label{fig:S_charge_density:bound_2.5_FT}
		\phantomcaption\label{fig:S_charge_density:bound_3.0_FT}
		\phantomcaption\label{fig:S_charge_density:bound_3.5_FT}
		\phantomcaption\label{fig:S_charge_density:full_0.2_real}
		\phantomcaption\label{fig:S_charge_density:full_0.6_real}
		\phantomcaption\label{fig:S_charge_density:full_1.2_real}
		\phantomcaption\label{fig:S_charge_density:full_0.2_FT}
		\phantomcaption\label{fig:S_charge_density:full_0.6_FT}
		\phantomcaption\label{fig:S_charge_density:full_1.2_FT}
	\end{subcaptiongroup}
	\caption{
		Numerical simulations of the spatial distributions of the charge density around a vacancy site.
		\subref{fig:S_charge_density:bound_2.5_real}--\subref{fig:S_charge_density:bound_3.5_real} and \subref{fig:S_charge_density:full_0.2_real}--\subref{fig:S_charge_density:full_1.2_real} show the charge density at each site, and \subref{fig:S_charge_density:bound_2.5_FT}--\subref{fig:S_charge_density:bound_3.5_FT} and \subref{fig:S_charge_density:full_0.2_FT}--\subref{fig:S_charge_density:full_1.2_FT} show corresponding Fourier transforms in the first Brillouin zone.
		(a)--(f) Simulations for the bound-flux sector.
		$t_{bc}=1.82$, $\ell_0=2$, and $\chi_0=2.5$ for \subref{fig:S_charge_density:bound_2.5_real} and \subref{fig:S_charge_density:bound_2.5_FT}, $3.0$ for \subref{fig:S_charge_density:bound_3.0_real} and \subref{fig:S_charge_density:bound_3.0_FT}, and $3.5$ for \subref{fig:S_charge_density:bound_3.5_real} and \subref{fig:S_charge_density:bound_3.5_FT}.
		(g)--(l) Simulations for the full-flux sector.
		$t_{bc}=0.1$, $\ell_0=2$, and $\chi_0=0.2$ for \subref{fig:S_charge_density:full_0.2_real} and \subref{fig:S_charge_density:full_0.2_FT}, $0.6$ for \subref{fig:S_charge_density:full_0.6_real} and \subref{fig:S_charge_density:full_0.6_FT}, and $1.2$ for \subref{fig:S_charge_density:full_1.2_real} and \subref{fig:S_charge_density:full_1.2_FT}.
		The incommensurate oscillating patterns are seen around the vacancy.
		For all of these calculations, the system consists of 3200 sites.
	}\label{fig:S_charge_density}
\end{figure}

With the use of the model Hamiltonian $\mathcal{H}$, we calculate the spatial distribution of the charge density.
Although the system is a Mott insulator with no explicit charge degrees of freedom, the charge density is expressed in terms of spin correlation functions between sites forming a triangle~\cite{Bulaevskii08,Pereira20},
\begin{eqnarray}
	\delta n_j \sim e\sum_{j,k,l}\ab[
		\braket[1]{S^{\alpha}_j S^{\alpha}_k} +\braket[1]{S^{\alpha}_j S^{\alpha}_l} - 2 \braket[1]{S^{\alpha}_k S^{\alpha}_l}
	].\label{eq:cd}
\end{eqnarray}
To calculate Eq.~(\ref{eq:cd}), we need to specify the configuration of visons.
Since the \ac{STM} measurements were performed at temperatures sufficiently lower than the vison gap, it may be appropriate to assume the bound-flux sector which is the ground state of the Kitaev \ac{QSL} with a vacancy; \textit{i.e.} a vison exists only at the vacancy site.
However, on the other hand, bias voltages used for the experiment are larger than the Mott gap, which indicates that visons may be excited in the measurements.
Thus, we examine two distinct vison configurations; one is the bound-flux sector, and the other one is the full-flux sector where all of the hexagons in the honeycomb lattice are occupied by visons.
As will be shown below, we can reproduce the incommensurate oscillating patterns observed in the \ac{STM} measurements for both of these vison configurations, if we choose model parameters properly.
Also, we chose $\ell_0=2$ for the characteristic length of the damping factor in Eq.~(\ref{eq:hssc}).
We note that the period of the oscillation does not strongly depend on the value of $\ell_0$, though the amplitude of the oscillation is affected by it.
In Figs.~\ref{fig:S_charge_density:bound_2.5_real}, \ref{fig:S_charge_density:bound_3.0_real}, and \ref{fig:S_charge_density:bound_3.5_real}, we show the calculated results of the spatial distribution of the charge density $\delta n_j $ for the bound-flux sector, which exhibit spatially oscillating behaviors around the vacancy.
The incommensurate character of the oscillation is more clearly seen in the Fourier transform of the charge distribution shown in Figs.~\ref{fig:S_charge_density:bound_2.5_FT}, \ref{fig:S_charge_density:bound_3.0_FT}, and \ref{fig:S_charge_density:bound_3.5_FT}.
The calculated results for the full-flux sector are also shown in Figs.~\ref{fig:S_charge_density:full_0.2_real}--\ref{fig:S_charge_density:full_1.2_FT}.
For both of the bound-flux sector and the full-flux sector, we can find parameters for which the incommensurate peaks in the first Brillouin zone are in agreement with the experimental observations.
In the case of the bound-flux sector, we used the value of the $bc$-hopping parameter $t_{bc}=1.83$ which is considerably large.
Since this term arises from the $\Gamma^\prime$ term, and $t_{bc}=\Gamma^\prime$, one may expect that $t_{bc}$ can not exceed $K=1$ to stabilize the Kitaev \ac{QSL} state.
However, a lattice distortion generally occurs in the vicinity of the vacancy, which lowers lattice symmetry, and leads to the enhancement of non-Kitaev interactions.
Thus, it may be possible that $t_{bc}$ exceeds $K=1$ locally preserving the bulk Kitaev \ac{QSL} state.
The incommensurate oscillating patterns shown in Fig.~\ref{fig:S_charge_density} may remind us of the Friedel oscillation of electrons in metals.
However, there is an important difference between the oscillation in the Kitaev \ac{QSL} and the Friedel oscillation.
The conventional Friedel oscillation of electrons arises from the screening of a long-range Coulomb potential.
On the other hand, in the Kitaev \ac{QSL} state, the Majorana oscillation is caused by scatterings with the gauge $b$-Majoranas neighboring the vacancy site, described by Eq.~(\ref{eq:hnk}).
The Majorana Fermi surface shown in Figs.~\ref{fig:S_Majorana_Fermi_surface:1.5}, \ref{fig:S_Majorana_Fermi_surface:2.5}, and \ref{fig:S_Majorana_Fermi_surface:3.0} is slightly deformed by the existence of the vacancy via the hybridization with the gauge $b$-Majoranas, described by Eq.~(\ref{eq:hnk}).
The incommensurate wavenumbers found in Figs.~\ref{fig:S_charge_density:bound_2.5_FT}--\ref{fig:S_charge_density:bound_3.5_FT} and \ref{fig:S_charge_density:full_0.2_FT}--\ref{fig:S_charge_density:full_1.2_FT} are mainly determined by the differences of two Fermi wavenumbers on the deformed Majorana Fermi surface.
We show the comparison between the experimental observations and the numerical simulation in Fig.~\ref{fig:FT_comparison}.
This scenario of the Majorana oscillation provides a promising explanation for the observation of the \ac{STM} measurements.

Now we discuss more precisely the comparison between the theoretical results and the experimental observations.
The incommensurate peaks observed in the \ac{STM} measurements do not depend on bias voltages, at least in the plus bias region $V>0$ or the minus bias region $V<0$, as shown in Fig.~\ref{fig:map:dispersion}.
This behavior is quite different from quasiparticle interference patterns usually observed in metals.
It is natural to expect that the bias independence may be attributed to the existence of the definite incommensurate Fermi wavenumbers of Majorana particles.
In the above scenario, the Fermi wavenumbers of itinerant Majorana particles depend on $\chi_0$ in Eq.~(\ref{eq:hssc}).
The coefficient $\chi_0$ can be regarded as a mean value of the scalar spin chirality, which is determined by the interplay between exchange interactions and the tunnel current which induces circular orbital currents associated with the spin chirality.
The bias-independence of the oscillating period observed in the \ac{STM} measurements implies that the mean value $\chi_0$ is mainly determined by the exchange interactions among relevant spins, and the tunnel current merely acts as a trigger to induce a metastable state with spin chirality configuration.
On the other hand, the periods of the spatial oscillation are slightly different between the positive bias \SI[explicit-sign=+]{1}{V} and the minus bias \SI{-1}{V}, as shown in Fig.~\ref{fig:oscillation_FT} and Fig.~\ref{fig:map}.
A possible origin of this asymmetry is the local change of the Kitaev interaction caused by the bias potential. Since the energy difference between the $j=1/2$ and $j=3/2$ states is \SI{\sim 0.15}{eV} for \ce{\alpha-RuCl3}~\cite{Kim15}, the bias \SI{\pm 1}{V} may affect virtual processes associated with the Kitaev interaction, which arises from the transition between the $j=1/2$ state and the $j=3/2$ state.
That is, the negative bias can result in virtual states with extra holes, while the positive bias does not generate such virtual processes.
In fact, with the virtual processes with extra holes, the Kitaev interaction increases as described by $\dfrac{t^2J_\mathrm{H}}{U^2}+\dfrac{t^2J_\mathrm{H}}{U^2}\ab(\dfrac{t_\mathrm{S}}{U})^2$.
Here, the first term represents the conventional Kitaev interaction, and the second term is an enhancement in the negative bias.
$t$ is the electron hopping amplitude between adjacent sites, $U$ is the Hubbard onsite repulsion, $J_\mathrm{H}$ is the Hund's coupling, and $t_\mathrm{S}$ is the transition amplitude for the process where one electron is removed at the negative bias.

We confirm this prediction by performing model calculations.
In order to qualitatively investigate possible changes in the Kitaev interaction under the \ac{STM} setup, we consider the following electronic toy model for $t_{2g}$ electrons on the hexagonal lattice:
\begin{align}
	\mathcal{H} = \mathcal{H}_{\mathrm{SOC}}+\mathcal{H}_{U}+\mathcal{H}_{\mathrm{kin}}+\mathcal{H}_{\mathrm{pot}}.
	\label{eq:hubbard}
\end{align}
The first term in Eq.~(\ref{eq:hubbard}) describes spin-orbit coupling:
\begin{align}
	\mathcal{H}_{\mathrm{SOC}}
	= \dfrac{\lambda}{2}\sum_{i}\bm{d}^{\dagger}_{i}\ab(\hat{\bm{L}}\cdot\hat{\bm{\sigma}})\bm{d}^{\phantom{\dagger}}_{i}
	= \dfrac{\lambda}{2}\sum_{i}\bm{d}^{\dagger}_{i}
		\begin{pmatrix}
			0 & i\hat{\sigma}^{z} & -i\hat{\sigma}^{y} \\
			-i\hat{\sigma}^{z} & 0 & i\hat{\sigma}^{x} \\
			i\hat{\sigma}^{y}  & -i\hat{\sigma}^{x} & 0
		\end{pmatrix}
		\bm{d}_{i}^{\phantom{\dagger}},
\end{align}
where $d^{\dagger\sigma}_{im}$ ($d^{\phantom{\dagger}\sigma}_{im}$) is the creation (annihilation) operator of an electron with spin $\sigma$ at site $i$ and orbital $m$ running over the $t_{2g}$ states (in the order of $yz$, $zx$, and $xy$ orbitals), or $\bm{d}^{\dagger}_{i}=(d^{\dagger\uparrow}_{i,yz}\,d^{\dagger\downarrow}_{i,yz}\,d^{\dagger\uparrow}_{i,zx}\,d^{\dagger\downarrow}_{i,zx}\,d^{\dagger\uparrow}_{i,xy}\,d^{\dagger\downarrow}_{i,xy})$, $\bm{L}$ is the orbital angular momentum in the basis of the $t_{2g}$ states, and $\hat{\bm{\sigma}}$ is a vector of Pauli matrices. The second term is the Hubbard-Kanamori Hamiltonian describing the on-site Coulomb interactions:
\begin{align}
	\mathcal{H}_{U}
	&= U\sum_{i,m}n_{im}^{\uparrow}n_{im}^{\downarrow}
	+ U'\sum_{i,m\ne m'}n_{im}^{\uparrow}n_{im'}^{\downarrow}
	+ (U'-J_{\mathrm{H}})\sum_{i,m<m',\sigma}n_{im}^{\sigma}n_{im'}^{\sigma}\nonumber\\
	& +J_{\mathrm{H}}\sum_{i,m\ne m'}d^{\dagger\uparrow}_{im}d^{\dagger\downarrow}_{im'}d^{\phantom{\dagger}\downarrow}_{im}d^{\phantom{\dagger}\uparrow}_{im'}
	+ J_{\mathrm{H}}\sum_{i,m\ne m'}d^{\dagger\uparrow}_{im}d^{\dagger\downarrow}_{im}d^{\phantom{\dagger}\downarrow}_{im'}d^{\phantom{\dagger}\uparrow}_{im'},
\end{align}
where $n_{im}^{\sigma}=d^{\dagger}_{im}d^{\phantom{\dagger}}_{im}$ is the density operator, $U$ is the intraorbital Coulomb interaction, $J_{\mathrm{H}}$ is the Hund's coupling, and the interorbital Coulomb interaction is taken as $U'=U-2J_{\mathrm{H}}$. The third term is kinetic energy given by hopping parameters:
\begin{align}
	\mathcal{H}_{\mathrm{kin}} = \sum_{\substack{ij,mm' \\ \sigma}}t_{ij}^{mm'}d^{\dagger\sigma}_{im}d^{\phantom{\dagger}\sigma}_{jm'}.
\end{align}
\noindent Hopping parameters $\hat{T}_{ij}=(t_{ij}^{mm'})$ are chosen depending on the bond type so that the only non-zero term of the corresponding spin model is the Kitaev interaction:\cite{PhysRevB.93.214431}
\begin{align}
	\hat{T}_{\mathrm{X}} =
	\begin{pmatrix}
		0 & 0 & 0 \\
		0 & 0 & t \\
		0 & t & 0 \\
	\end{pmatrix}
	\qquad
	\hat{T}_{\mathrm{Y}} =
	\begin{pmatrix}
		0 & 0 & t \\
		0 & 0 & 0 \\
		t & 0 & 0 \\
	\end{pmatrix}
	\qquad
	\hat{T}_{\mathrm{Z}} =
	\begin{pmatrix}
		0 & t & 0 \\
		t & 0 & 0 \\
		0 & 0 & 0 \\
	\end{pmatrix}
\end{align}
The last term is the on-site potential that accounts for the local change in the number of electrons under the \ac{STM} setup:
\begin{align}
	\mathcal{H}_{\mathrm{pot}} = \varepsilon\sum_{m,\sigma} n_{im}^{\sigma}.
\end{align}
One should stress that $\varepsilon$ does not directly represent the bias voltage, and is introduced to simulate an extra hole or an electron added to the system in the case with the applied voltage.

\begin{figure}[t]
	\centering
	\includegraphics[width=0.3\textwidth]{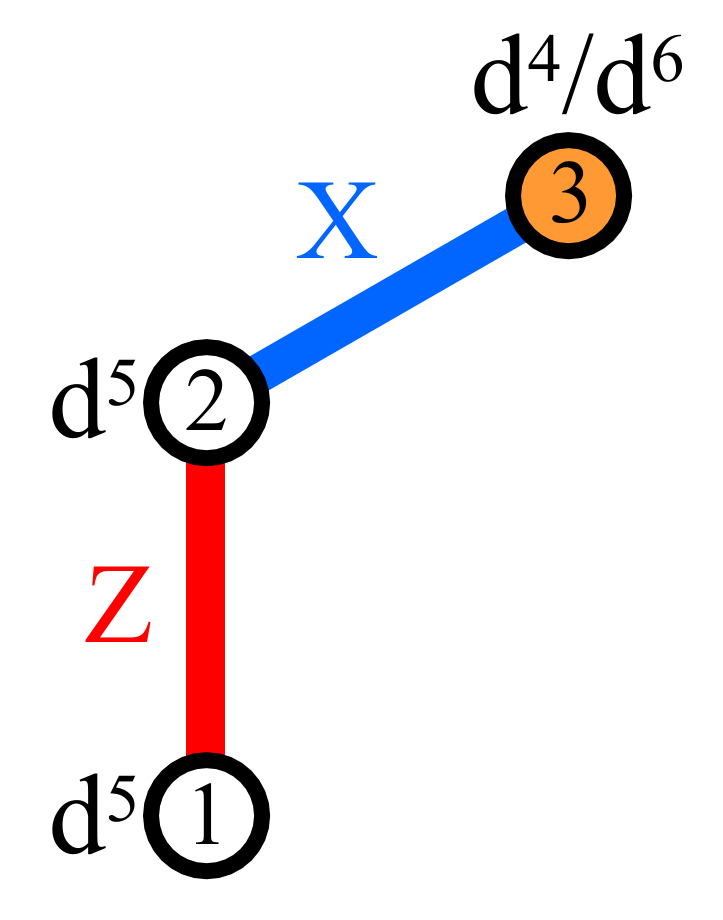}
	\caption{Schematic view of the toy Hubbard model.}
	\label{fig:model}
\end{figure}

Calculations are carried out using the following model parameters: $U=3.66$, $J_{\mathrm{H}}=0.31$, $U'=U-2J_{\mathrm{H}}$, $\lambda=0.118$, $t=0.1$, where the values of Coulomb interactions and spin-orbit coupling are obtained from density functional theory for monolayer \ce{RuCl3}.
We would like to note that we do not consider intersite Coulomb interactions, crystal-field splitting, and realistic hopping parameters on the level of this toy model.

In the ground state, each site accommodates five electrons ($d^{5}$ configuration) resulting in a $J=1/2$ state.
Upon solving the Hubbard model for two and three sites (with $N=$ 10 and 15 electrons, respectively), the low-lying states can be projected onto the spin model formulated in the basis of $J=1/2$ states ($\ket{\sigma,\sigma}$ or $\ket{\sigma,\sigma^\prime,\sigma''}$ with $\sigma$, $\sigma'$, $\sigma''=\uparrow,\downarrow$)~\cite{PhysRevLett.73.1873,PhysRevB.65.104508,PhysRevB.70.104424}.
With the hopping parameters considered above, the only non-zero term is calculated to be the Kitaev interaction: $K=-0.000872$ and $-0.000864$ for two and three sites, respectively. Note that a small change comes from higher-order hopping processes~\cite{PhysRevB.93.214431}.

In \ac{STM}, an extra electron or a hole is injected into the system depending on whether a positive or negative voltage is applied, resulting in a local change of the electronic configuration as $d^{6}$ or $d^{4}$, respectively.
Taking into account that spin-orbit coupling splits the ninefold degeneracy of four electrons in the $t_{2g}$ subspace, both $d^{6}$ and $d^{4}$ states are non-magnetic with the total angular momentum $J=0$.
Given a supercell with three sites, we further assume that an electron is placed or removed at site 3, as shown in Fig.~\ref{fig:model}.
As a technical sidenote, choosing a sufficiently large on-site potential $\varepsilon>0$ ($\varepsilon<0$) allows us to fix a $d^{5}$-$d^{5}$-$d^{4}$ ($d^{5}$-$d^{5}$-$d^{6}$) configuration in the ground state manifold and lift the $d^{5}$-$d^{4}$-$d^{5}$ ($d^{5}$-$d^{6}$-$d^{5}$) and $d^{4}$-$d^{5}$-$d^{5}$ ($d^{6}$-$d^{5}$-$d^{5}$) states to higher energies, which are regarded equivalent otherwise for $\varepsilon=0$.
Upon solving the three-site Hubbard model with $N=14$ and 16 electrons, the low-energy spectrum can be mapped onto the $\ket{\sigma,\sigma',0}$ basis states with $\sigma$, $\sigma'=\uparrow,\downarrow$ giving the spin model parameters between sites 1 and 2.
From the results presented in Table~\ref{table:kitaev}, one can see that the Kitaev interaction can be modified due to different hopping pathways which appear when injecting or removing an electron in close vicinity to a given bond.

\begin{table}[]
	\caption{Kitaev interaction $K$ between sites 1 and 2 calculated for different choices of the on-site potential $\varepsilon$ in the model with three sites and an extra hole ($d^{5}-d^{5}-d^{4}$) or an electron ($d^{5}-d^{5}-d^{6}$) placed at site 3.}
	\begin{tabular}{c|c||c|c}
		\hline
		\multicolumn{2}{c||}{$d^{5}$-$d^{5}$-$d^{4}$} & \multicolumn{2}{c}{$d^{5}$-$d^{5}$-$d^{6}$} \\
		\hline
		$\varepsilon$ & $K$ & $\varepsilon$ & $K$ \\
		\hline
		1.0  & $-0.000892$ & $-1.0$ & $-0.000845$ \\
		0.9  & $-0.000898$ & $-0.9$ & $-0.000843$ \\
		0.8  & $-0.000908$ & $-0.8$ & $-0.000839$ \\
		0.7  & $-0.000920$ & $-0.7$ & $-0.000833$ \\
		0.6  & $-0.000940$ & $-0.6$ & $-0.000827$ \\
		0.5  & $-0.000970$ & $-0.5$ & $-0.000818$ \\
		\hline
	\end{tabular}
	\label{table:kitaev}
\end{table}

\end{document}